\documentclass[fleqn,usenatbib]{mnras}

\usepackage{newtxtext,newtxmath}

\usepackage[T1]{fontenc}

\DeclareRobustCommand{\VAN}[3]{#2}
\let\VANthebibliography\thebibliography
\def\thebibliography{\DeclareRobustCommand{\VAN}[3]{##3}\VANthebibliography}

\usepackage{graphicx}	
\usepackage{amsmath}	
\usepackage{amssymb}	
\usepackage{bbold}
\usepackage{multicol}        
\usepackage{bm}		
\usepackage{pdflscape}	
\usepackage{multirow}
\usepackage{booktabs}
\usepackage{graphicx}
\usepackage{array}
\usepackage{float}
\usepackage{xspace} 
\usepackage{subfiles} 
\usepackage{makecell}
\usepackage{comment}

\newcommand{\inv}{^{\raisebox{.2ex}{$\scriptscriptstyle-1$}}}
\newcommand{\mat}[1]{\mathbf{#1}}

\newcommand{\Planck}{\emph{Planck}\xspace}
\newcommand{\WMAP}{\emph{WMAP}\xspace}

\newcommand{\madam}{\texttt{Madam}\xspace}
\newcommand{\SRone}{\texttt{SRoll1}\xspace}
\newcommand{\SRtwo}{\texttt{SRoll2}\xspace}

\newcommand{\glass}{\textsc{glass}\xspace}
\newcommand{\momento}{\texttt{momento}\xspace}
\newcommand{\pixlike}{\texttt{pixLike}\xspace}

\newcommand{\LCDM}{$\Lambda$CDM\xspace}

\def\ltsima{$\; \buildrel < \over \sim \;$}
\def\gtsima{$\; \buildrel > \over \sim \;$}
\def\simlt{\lower.5ex\hbox{\ltsima}}
\def\simgt{\lower.5ex\hbox{\gtsima}}

\newcommand{\bsync}{\beta_{\rm sync}}

\newcommand{\bdust}{\beta_{\rm dust}}

\providecommand{\sorthelp}[1]{}
\usepackage[dvipsnames]{xcolor}

\newcolumntype{N}{>{\centering\arraybackslash}m{.5in}}
\newcolumntype{G}{>{\centering\arraybackslash}m{2in}}

\defcitealias{2016A&A...596A.107P}{PSRoll1}
\defcitealias{Aghanim:2018eyx}{PCP18}
\defcitealias{2020A&A...641A...5P}{PLP19}
\defcitealias{2019A&A...629A..38D}{PSRoll2}
\defcitealias{EG19}{EG}
\defcitealias{Belsunce21}{B21}

\title[Testing polarised CMB foregrounds]{Testing for spectral index variations in polarised CMB foregrounds}

\author[de Belsunce, Gratton \& Efstathiou]{Roger de Belsunce,\thanks{E-mail: \href{mailto:rmvd2@cam.ac.uk}{rmvd2@cam.ac.uk}} Steven Gratton, and George Efstathiou
\\Kavli Institute for Cosmology \& Institute of Astronomy, University of Cambridge, Madingley Road, Cambridge CB3 OHA, United Kingdom}

\date{Accepted XXX. Received YYY; in original form ZZZ}

\pubyear{2022}

\begin{document}
\label{firstpage}
\pagerange{\pageref{firstpage}--\pageref{lastpage}}
\maketitle

\begin{abstract}
We present a Bayesian parametric component separation method for polarised microwave sky maps. We solve jointly for the primary cosmic microwave background (CMB) signal  and the main Galactic polarised foreground components. For the latter, we consider electron-synchrotron radiation and thermal dust emission, modelled in frequency as a power law  and a modified blackbody respectively. We account for inter-pixel correlations in the noise covariance matrices of the input maps and introduce a spatial correlation length in the prior matrices for the spectral indices $\beta$. We apply our method to  low-resolution polarised \Planck 2018 Low and High Frequency Instrument (LFI/HFI) data, including the \SRtwo re-processing of HFI data. We find evidence for spatial variation of the synchrotron spectral index, and no evidence for depolarisation of dust. Using the HFI \SRtwo maps, and applying wide priors on the spectral indices, we find  a mean polarised synchrotron spectral index over the unmasked sky of $\Bar{\beta}_{\rm sync}=-2.833\pm 0.620$. For polarised thermal dust emission, we obtain $\Bar{\beta}_{\rm dust}=1.429\pm 0.236$. Our method returns correlated uncertainties for all components of the sky model. Using our recovered CMB maps and associated uncertainties, we constrain the optical depth to reionization, $\tau$, using a cross-spectrum-based likelihood-approximation scheme (\momento) to be $\tau = 0.0598\pm 0.0059$. We confirm our findings using a pixel-based likelihood (\pixlike). In both cases, we obtain a result that is consistent with, albeit a fraction of a $\sigma$ higher than that found by subtracting spatially uniform foreground templates. While the latter method is sufficient for current polarisation data from \Planck, next-generation space-borne CMB experiments will need more powerful schemes such as the one presented here.
\end{abstract}




\section{Introduction} \label{sec:intro}
Measurements of the polarised cosmic microwave background (CMB)
anisotropies provide a unique opportunity to constrain early Universe
physics. However, Galactic polarised emission must be removed to high
accuracy to extract reliable information from the primordial
CMB. Disentangling the polarised CMB signal at large angular scales
from its contaminants, mainly  synchrotron emission 
and thermal dust emission, is crucial for
measurement of the reionization optical
depth $\tau$ and the detection of a signal from
primordial gravitational waves.

Scalar perturbations generated during inflation give a
divergence-like $E$-mode polarisation pattern in the CMB, whereas
tensor perturbations would produce a distinctive curl-like $B$-mode
polarisation signature, together with an $E$-mode pattern of
approximately equal amplitude \citep{Kamionkowski:1997,Seljak:1997}.
Measuring primordial $B$-modes would fix the energy scale of inflation
\citep{Starobinskii1983} and provide constraints on specific
inflationary models \citep[e.g.][]{Linde:2019}.  Gravitational lensing
of CMB $E$-modes by intervening matter generates a $B$-mode anisotropy
contributing an additional contaminant to the primordial B-mode
signal\footnote{This effect, however, is negligibly small on the
scales analysed in this present paper (multipoles $\ell \leq 30$).}
\citep[see][for a review]{Lewis2006}.

A key science goal of next-generation CMB experiments from space
\citep[LiteBIRD;][]{2020JLTP..199.1107S} and from the ground, such as
CMB-S4 \citep{CMBS4:2020} and the Simons Observatory
\citep[SO,][]{SO_2019} is to achieve a statistically significant
detection of primordial gravitational waves if the tensor-to-scalar
ratio $r$ is of order $10^{-3}$ \citep{SO_2019}. Polarised foregrounds
pose a major challenge for these ambitious projects. It is therefore
important to investigate Galactic foregrounds in as much detail as
possible using existing polarisation data.  In this paper we use
\Planck  data at large angular scales, which poses
additional challenges because of the low signal-to-noise of the \Planck polarisation maps 
and the presence of residual
instrumental systematics.

We adopt a Bayesian approach to polarized foreground removal. The aim
is to recover the polarised CMB anisotropies, together with a
parametric model of polarised Galactic foregrounds, using a
statistical framework that allows the foreground errors to be
propagated through to cosmological parameters.  We  assume that lower
frequencies ($\nu \simlt 70\ {\rm GHz}$) are dominated by  synchrotron emission
while higher frequencies ($\nu \simgt 100\ {\rm GHz}$) are dominated by thermal
dust emission. We approximate the frequency dependence of the polarised foregrounds as a power law for synchrotron and a modified blackbody for thermal dust. Aspects of the theoretical basis for the approach used in this
paper were originally developed in e.g.~\citet{Eriksen2006, Dunkley:2009fg, Stompor09} and \citet{Gratton2008}. We extend this framework by allowing for pixel-pixel correlations in the noise covariance 
matrices (NCM) and imposing priors on the variations of the 
spectral indices across the sky, effectively regularising noise 
in the maps.

For Low Frequency Instrument (LFI) \Planck data \citep{LFI_data:2018}, we assume that
polarised emission arises  primarily from synchrotron radiation emitted by
spiralling cosmic ray electrons in the Galactic magnetic field. Synchrotron
emission can be highly polarised, with polarisation fractions of
up to  75\% \citep{Ginzburg:1965}. The specific intensity, $I_\nu$, of
synchrotron emission has power law frequency dependence, 
$I_\nu \propto \nu^{\beta_s}$, with $\beta_s\sim -3$. The spectral index $\beta_s$ is
related to the energy distribution of the cosmic rays of cosmic ray electrons, $dN/dE \propto E^{-p}$, according to $\beta_s = -(p+3)/2$ \citep{Rybicki:1979}. There have been many studies of 
angular variations of he spectral index $\beta_s$. Radio frequency observations
in the frequency range  408 MHz to 10 GHz \citep{Davies:1996, Platania:2003,
  Bennett:2003} find values of $\beta_s$ in the range -2.8 to 3.2 with shallower
values towards the Galactic plane. This is consistent with models in which the
steepening of the spectral index is caused by energy dependent diffusion and radiative
losses as electrons propagate into the Galactic halo \citep[see e.g.][]{Dickinson_2018}.
There have also been analyses reporting steeper synchrotron spectral indices in the radio 
loops, perhaps reflecting an upper limit to the electron energy spectrum generated by shock acceleration  within the
loops \citep{Lawson:1987, Reich:1988, Borka:2007}. Galactic foreground emission in total intensity over the frequency range
$20-100$ GHz  is discussed in detail by \cite{Planck_low-frequency_foregrounds:2016} and is difficult
to interpret because of the complexities associated with disentangling synchrotron from free-free and anomalous microwave
emission. Polarised Galactic emission over  this frequency range, which is the subject of this paper, is dominated 
by synchrotron emission. However,  polarization maps from \Planck and the Wilkinson Microwave Anisotropy Probe \citep[\WMAP;][]{2013ApJS..208...19H} are noisy and affected by instrumental systematics.
Previous work has largely been limited to measuring the synchrotron spectral index $\beta_s$ leading to values
$\beta_s \approx -3$ (consistent with the synchrotron intensity spectral index) with some tentative evidence
of spectral index variations \citep{Page:2007, Fuskeland:2014, Vidal2015,Dickinson_2018, Svalheim:2020, Fuskeland:2021, Martire:2021}.

At frequencies $\simgt 217$ GHz covered by the \Planck 
High Frequency Instrument (HFI) \citep{HFI_data:2018, 2019A&A...629A..38D} thermal Galactic dust emission
dominates the foreground emission in intensity over most of the sky, 
with the cosmic infrared background (CIB) dominating at high Galactic latitudes.
Galactic dust emission has been discussed  by the \Planck
collaboration \citep{Planck_dust:2014, Planck_dust:2015}. Using map-based statistics,
these  papers find that the spectral energy distribution (SED) of Galactic dust emission in intensity is well approximated by a 
modified black body distribution with a spectral index $\beta_d = 1.59 \pm 0.12$, 
and dust temperature $T_d = 20.3 \pm 1.3 {\rm K}$. A power-spectrum based analysis,
which allows accurate separation of Galactic dust emission from the  CIB and  primordial
CMB  leads to consistent results of $\beta_d = 1.49 \pm 0.05$ and $T_d = 22.7 \pm 2.8 {\rm K}$
\citep{EG19}. There is no evidence at the \Planck\ frequencies for angular variations of the 
diffuse Galactic dust emission in intensity.

Since the highest polarised \Planck channel is at 353 GHz,  there is  a more
restricted frequency range available with which to study  Galactic dust emission  in polarization
compared to intensity.  Nevertheless, the polarised dust SED
is consistent with a modified black body distribution with the same values of $\beta_d$ and
$T_d$ as in intensity \citep{Planck_dust:2015}. Variations in the dust temperature and orientation of
dust grains with respect to the Galactic magnetic field can lead to line-of-sight differences in the polarised 
dust SED \citep{Tassis:2015}. An analysis of the \Planck BB power spectrum suggested a decorrelation of polarised
dust spectrum between $217$ and $353$ GHz \citep{2017A&A...599A..51P}.  However a more detailed analysis of the \Planck maps, using more robust statistical techniques, showed no evidence for any decorrelation \citep{2020A&A...641A..11P}. More
recently \cite{Pelgrims:2021} have used a combination of \Planck\ polarization data and measurements of 
neutral hydrogen (HI) velocity components to find that lines of sight with multiple HI components display 
greater depolarisation compared to other lines of sight.

Many component separation methods have been developed to disentangle
the primary CMB signal from foregrounds \citep[see e.g.][for
  examples]{Planck2020_compsep}. The methods can be roughly split into
three categories: template fitting, parametric fitting methods and
multi-frequency internal linear combination schemes.  The analysis of
\Planck polarization maps has been based on subtracting a low
frequency synchrotron template and a high frequency Galactic dust
template \citep{2016A&A...596A.107P, 2020A&A...641A...5P, Pagano:2020,
Belsunce21}.  Bayesian parametric fitting methods, such as the
Markov Chain Monte Carlo algorithm \texttt{Commander}
\citep{Eriksen:2004, Eriksen2008}, have been applied successfully to
disentangle multipole foreground components in intensity \citep[see
  e.g.][]{Planck2020_compsep}. However these methods have been much
less successful when applied to large-scale polarization because of
the higher noise levels (leading to convergence problems) and the
contribution of instrumental systematics. For an early application
of pixel-by-pixel parametric foreground fitting to WMAP polarization
data see \cite{Page:2007}. In this paper, we develop and apply stable
Bayesian parametric methods to the removal of Galactic foregrounds
from the \Planck polarization maps.

This paper is organised as follows: in
Sec.~\ref{sec:parametric_fitting} we introduce our Bayesian parametric
fitting method and in Sec.~\ref{sec:likelihoods} we present our likelihoods. In Sec.~\ref{sec:data} we discuss the \Planck data
that we use and the pixel-pixel noise covariance matrices required for
the Bayesian component separation method. In Sec.~\ref{sec:results} we test
the parametric foreground fitting approach on \Planck data and use two
likelihoods to quantify its effect on the optical
depth to reionization $\tau$ parameter. Sec.~\ref{sec:conclusion}
presents our conclusions.

\section{Methods} \label{sec:parametric_fitting}

Following the formalism discussed in  e.g.~\citet{Eriksen2008, Stompor09, Gratton_Grenoble, Dunkley:2009fg, Armitage-Caplan2011} and \citet{Gratton2008}, we apply a
Bayesian parameter estimation framework in which we infer the posterior density $\mathcal{P}(\bm{\theta} \vert \mathbf{d}, \mathcal{M})$ of a set of foreground and signal parameters $\bm{\theta}$ describing a model $\mathcal{M}$ from a realisation of data $\mathbf{d}$:
\begin{equation}
    \label{eq:Bayes_theorem}
   \mathcal{P}(\mathbf{\theta}|\mathbf{d}, \mathcal{M}) = \frac{\mathcal{P}(\mathbf{d}|\mathbf{\theta}, \mathcal{M}) \mathcal{P} (\mathbf{\theta}| \mathcal{M})}{\mathcal{P}(\mathbf{d}| \mathcal{M})}\ ,
\end{equation}
where $\mathcal{P}(\mathbf{\theta}|\mathbf{d}, \mathcal{M})$ is the posterior, $\mathcal{P}(\mathbf{d}|\mathbf{\theta}, \mathcal{M})$ the likelihood, $\mathcal{P} (\mathbf{\theta}| \mathcal{M})$ the prior and $\mathcal{P}(\mathbf{d}| \mathcal{M})$ the evidence.

The \Planck satellite provides $I$, $Q$, and $U$ maps, each containing $N_{\rm pix}$ pixels,  in $N_{\nu}$ frequency bands. We assume Gaussian noise that is uncorrelated between frequency channels but may have inter-pixel correlations. Fitting parametric foreground models is computationally intensive at full \Planck resolution. Since in this paper we are interested in polarized foregrounds at large angular scales, we degrade the \Planck LFI and HFI maps to a resolution of $N_{\rm side}=16$ containing $N_{\rm pix} = 12 \cdot N_{\rm side}^2=3072$ pixels following the procedure described in \citet[][hereafter \citetalias{Belsunce21}]{Belsunce21}.

\subsection{Bayesian parametric foreground estimation} \label{sec:Likelihood_sec}
We assume that the observed sky can be modelled as
\begin{equation} \label{eq:model1}
    \mat d = \mat{Am} + \mat n \ ,
\end{equation}
where $\mat d$ are the observed (masked) sky maps in antenna temperature, $\mat{Am}$ is the sky signal which consists of a frequency-scaling matrix $\mat A$ and model amplitudes $\mat m$, and $\mat{n}$ is the instrumental noise. 

The parametric model for the sky signal is composed of three components: the primordial CMB ($c$), electron-synchrotron emission ($s$), and thermal dust radiation ($d$). Foregrounds are taken to be independent of the CMB and we model the contaminants, namely electron synchrotron and thermal dust, as a power law and a modified blackbody respectively, see e.g. \citet{Planck_diffuse_LFI:2016, Planck2020_compsep}.  Given Eq.\ \eqref{eq:model1}, we disentangle the CMB and the foregrounds by jointly estimating the posterior distribution $\mathcal{P}(\mat{A},\mat{m}\vert \mat{d})$ of the scaling matrix and the model amplitudes given the observed data.

The position-dependent scaling matrix $\mat A$ describes the frequency dependence of the CMB and foreground signals. The first columns of $\mat A$ consist of repeated block-diagonal identity matrices $\mathbb{1}$, weighting the primordial CMB, and the following columns are populated with factors weighting the foreground channels. The vector $\mat m$ consists of $N_{\rm var}$ stacked amplitudes of the model components: $\mat m_i$ with $i=\{c, s, d\}$. The sky signal can thus be written in thermodynamic temperature as the sum over the different components
\begin{align} \label{eq:model_fg}
    \mat{Am} &= \mat{m}_{\rm c}
    + \mat{m}_{s} \left(\frac{\nu}{\nu^{s}_0}\right)^{\pmb{\boldsymbol{\beta}}_{s}} \frac{g(x_0)}{g(x)} 
    + \mat{m}_{d} \left(\frac{\nu}{\nu^{d}_0}\right)^{\pmb{\boldsymbol{\beta}}_{d}+1} \frac{g(x_0)}{g(x)}\frac{f(y_0^{d})}{f(y)}\ ,
\end{align}
where $\nu_0^{i}$ ($i\in \{s,d\}$) are the pivot frequencies assigned to the channels assumed to best trace synchrotron and dust, respectively. The function $g(x)$ relates thermodynamic to brightness temperature and $f(y)$ is a frequency-dependent function accounting for a modified blackbody. The arguments are $x=h\nu / k_B T$ and $y=h\nu / k_B T_d$, with $k_B$ the Boltzmann, $h$ the Planck constant and $T$ ($T_d$) the CMB (dust) temperature\footnote{For the exact derivation and numerical values of the conversion factors, see appendix \ref{sec:appendix_spec_idx}.}. 

We introduce a physically motivated prior term for the spectral indices to vary continuously across the sky. Following the procedure outlined in \citet{Crowe_thesis:2013}, we therefore introduce a prior for the spectral indices incorporating a spatial correlation angle $\theta_{\rm corr}$, which adds non-zero off-diagonal entries to the prior:
\begin{equation} \label{eq:beta_cov1}
    \mat{C}_{\boldsymbol{\beta}} = \sigma(\boldsymbol{\beta}_i)\sigma(\boldsymbol{\beta}_j) \left[ \exp \left(-\frac{\theta_{ij}^2}{2\theta_{\rm corr}^2}\right) \right] \ .
\end{equation}
Here the angular separation\footnote{Taking the limit of a very large correlation angle reduces the scheme to a spatially uniform template cleaning procedure.} between the pixels across a sky map is given by $\theta_{ij}$, and $\sigma(\boldsymbol{\beta})^2$ is the variance per spectral index. The negative log-prior term thus reads 
\begin{equation}\label{eq:beta_cov2}
    \frac{1}{2}(\boldsymbol{\beta} - \Bar{\boldsymbol{\beta}})^\top \mat{C}\inv_{\boldsymbol{\beta}} (\boldsymbol{\beta} - \Bar{\boldsymbol{\beta}}) \ ,
\end{equation}
where $\boldsymbol{\beta}$ is a stacked vector of the spectral indices $(\boldsymbol{\beta}_s,\boldsymbol{\beta}_d)$. 

To determine the best solution, we minimise the negative log-likelihood ($\mathcal{S} \equiv -\ln\mathcal{P}$):
\begin{align} \label{eq:action}
    \mathcal{S} & = \frac{1}{2} \left[(\mat d-\mat{Am})^{\top} \mat N\inv (\mat d-\mat{A}\mat{m}) +(\boldsymbol{\beta} - \boldsymbol{\Bar{\beta}})^\top\mat{C}_{\boldsymbol{\beta}}\inv (\boldsymbol{\beta} - \boldsymbol{\Bar{\beta}}) \right]\ ,
\end{align}
where $\mat N$ is the noise covariance matrix (NCM) which we will introduce in Sec.\ \ref{sec:NCM}. The prior term acts as a penalty function on spectral index variations and effectively regularises scatter in the recovered maps. We solve this minimisation problem partly analytically and partly numerically. Uncertainties in our solution are estimated as the inverse of the curvature matrix of Eq.\ \eqref{eq:action}.

\subsubsection{Implementation details}
We sample the spectral indices jointly for $N_{\rm sp}=2$ \Planck polarisation maps, namely $Q$ and $U$, with $N_{\rm pix}=3072$ pixels in $N_{\nu}=6$ frequency channels: For LFI, we use 30 and 70 GHz and for HFI 100, 143, 217, and 353 GHz. We set the synchrotron pivot frequency to 30 GHz and the dust one to 353 GHz. We neglect noise correlations between frequency channels, using block-diagonal noise matrices in frequency space. The consideration of inter pixel noise covariance and correlated spectral index priors renders the minimisation problem computationally challenging\footnote{The presented approach is still much less expensive than Gibbs sampling approaches, see e.g.~\citep{Eriksen2008, Armitage-Caplan2011}. For low-resolution polarisation \Planck data with pixel-pixel correlations, the foreground cleaning takes a few CPU-hours at $N_{\rm side}=16$.}. The model degrees of freedom are the masked sky maps $\mat m = (\mat{m}_c^{Q,U}, \mat{m}_s^{Q,U}, \mat{m}_d^{Q,U})$ and $\boldsymbol{\beta} = (\boldsymbol{\beta}_s, \boldsymbol{\beta}_d)$, labelled collectively as $\mat x$, and our scheme requires inverting square matrices of corresponding size.

The minimisation is performed by a multi-dimensional pseudo-Newton-Raphson (NR) method \citep{NR_Press2007} following a two-step procedure. First, for given spectral indices, we analytically compute the amplitudes
\begin{equation} \label{eq:chi2_first_d}
    \mathbf{m} = (\mathbf{A}^{\top} \mathbf{N}\inv \mathbf{A})\inv\mathbf{A}^{\top}\mathbf{N}\inv \mathbf{d} \ , 
\end{equation}
that minimize Eq.\ \eqref{eq:action}. Second, we compute first and second derivatives of Eq.\ \eqref{eq:action} for all parameters and evaluate these at the assumed $\beta$'s and the calculated amplitudes. Using these derivatives, we take a step in parameter space $\Delta\mathbf{x} = -\gamma \mat{H}\inv \cdot \nabla \mathcal{S}$, where $H$ is the Hessian of the likelihood at the current trial point, and $\nabla \mathcal{S}$ the vector of first derivatives of Eq.\ \eqref{eq:action}. We iterate until convergence is achieved. The parameter $\gamma \in \left[ 0,1\right]$ is adaptively varied during the process to speed up convergence.  

For an initial starting point, we chose to use the 30 GHz maps for the synchrotron amplitudes, the 100 GHz maps for the CMB, and 353 GHz maps for the thermal dust amplitudes. For the spectral index maps, we chose to start at the peak of the assumed prior, to be discussed below. 

In our analyses we choose the mask presented in Fig.~\ref{fig:masks} which retains a fraction of $f=0.52$ of the sky (effects of mask choice are discussed in appendix \ref{app:mask_dependence}). For some choices of broad spectral priors, the scheme did not immediately converge. In these situations, we first solved with a sufficiently tight prior to achieve convergence. Then we gradually increased the prior towards the desired value, and using the previous solution as the starting point for the next one. Newton-Raphson-type methods are not guaranteed to converge and we saw examples of this phenomenon for a small number of pixels  ($\sim 10$) at the boundary of our original mask. We prevented this issue  from occurring in our analyses by masking these pixels together with their neighbours (for a total of $\sim 50$ additionally masked pixels).

\subsubsection{Noise and prior correlations} \label{sec:correlations}
To explore the sensitivity of our scheme to inter pixel correlations, we test three different cases. First, we take the noise covariance matrix (NCM) $\mat N$ and prior matrix $\mat C_{\boldsymbol{\beta}}$ to be diagonal; we denote this `diagNCM + diagPrior'. This simplification renders the problem computationally trivial as the minimisation can be carried out pixel by pixel. Second, we allow for pixel-pixel correlations in the NCM while keeping the prior matrix diagonal; we call this 'corrNCM + diagPrior'. Third, we allow for inter pixel correlations in both the NCM and the prior matrix; we denote this 'corrNCM + corrPrior'.

The resulting spectral index maps naturally depend on the spectral index priors because the data is not very constraining. Therefore, it is critical to choose physically motivated priors\footnote{One can easily analytically quantify the effect of a tight Gaussian prior combined with a large uncertainty from the data. Since the data is under-constraining the problem, the tight prior will lead to spectral index maps centred around the input prior value with only very low scatter. Adopting priors that are too tight can lead to results which are highly misleading.}. We introduce wide priors following the form of Eqs.\ \eqref{eq:beta_cov1} and \eqref{eq:beta_cov2}. These are centred around mean values of previous spectral index measurements from \citet{Planck2020_compsep} and radio frequency observations. Hence, we use $\bsync=-3.0$ with a prior width of $\sigma(\bsync)=2.0$, and $\bdust=1.5$ with a prior width of $\sigma(\bdust)=1.0$, for synchrotron and thermal dust respectively.

Significant variations of the spectral indices on small scales are unlikely because of the large interaction length of cosmic rays measured with gamma ray observations. Together with evidence for spatially varying spectral indices measured at 400 MHz - 10 GHz from radio frequency observations we are confident that spectral index variations occur at scales no smaller than $5-10^{\circ}$. Therefore, we choose an angular correlation length for both foreground components of $\theta_{\rm corr} = 5^{\circ}$.

\subsection{Likelihoods} \label{sec:likelihoods}
In the following, we briefly introduce and compare the two likelihoods used to assess the performance of the foreground cleaning scheme for cosmological parameter estimation. We do this by constraining the optical depth to reionization parameter $\tau$ from \Planck polarisation maps and compare our results to the template-cleaned analysis presented in \citetalias{Belsunce21}. 
In Sec.~\ref{sec:Likelihood_GLASS} we present a cross-spectrum-based semi-analytic likelihood approximation scheme, called \momento. In Sec.~\ref{sec:Likelihood_pixlike} we present a polarisation-only pixel-based likelihood. 

\subsubsection{Low multipole likelihood-approximation scheme} \label{sec:Likelihood_GLASS}
We present a cross-spectrum-based likelihood-approximation scheme (\momento) that is based on the  'General Likelihood Approximate Solution Scheme' \citep[\glass;][]{2017arXiv170808479G}. \glass allows to approximate the sampling distribution by computing moments and cumulants of the data. This is relevant in situations where the exact likelihood is unknown or computationally too expensive to compute. \momento was extensively tested on simulations and data in the context of low-$\ell$ CMB analysis in \citetalias{Belsunce21}. In practice, the moments can be either computed analytically or through forward simulations, to compute the gradient (denoted by $_{,a}$) of the log-likelihood:
\begin{equation}
    -(\log\mathcal{L})_{,a} = - (X - \langle X^\top \rangle)\langle \langle XX^\top \rangle \rangle \inv \langle X \rangle_{,a} \ .\label{eq:dsda}
\end{equation}
This procedure is generally applicable to a multitude of statistics $x^i,\ i \in \mathopen[1,\dots, n\mathclose]$, and $X$ denotes a vector of powers of varying degree of these statistics.
The $x^i$ ideally should be chosen to be as close to summary statistics as possible, e.g.\ elements of the power spectrum in the present case. It is also important to be able to obtain the required moments of the $x^i$ as a function of the model either analytically or with forward simulations.

For \momento we choose quadratic cross-spectra (QCS), compressing the CMB data down to a set of multipoles, here $\ell=2-29$. A fiducial power spectrum $C_\ell^{\mathrm{fid}}$ is assumed for the purpose of computing the QCS. Subsequently, for each evaluation of the likelihood the following actions are taken by \momento:
\begin{enumerate}
\item take input set of theory $C_\ell$'s;
\item compute the residual between fiducial and theory spectra $\Delta C_\ell = C_\ell-C_\ell^{\mathrm{fid}}$;
\item perform a path integral for the change of the log-likelihood in parameter space from $C_\ell^{\mathrm{fid}}$ to $C_\ell$. We choose the path $C_\ell^{\mathrm{fid}}  + \alpha \Delta C_\ell$ in QCS space, where $\alpha \in \left[ 0,1 \right]$. This requires computing the gradients of the log-likelihood, given in Eq.\ \eqref{eq:dsda}, along this path.
\end{enumerate}
To apply \momento, pixel-pixel covariance matrices are required to  compute moments of the QCS of the input maps. In \citetalias{Belsunce21}, we used NCMs of the frequency maps to compute $\tau$. In the present paper, \momento uses the CMB `block' of the inverse of the curvature covariance matrix which incorporates the uncertainties from the foreground cleaning scheme. 

In Sec.~\ref{sec:tau_sec} we will show results for $\tau$ using \momento on template-cleaned and CMB maps recovered from the presented component separation scheme. In the present paper, we compute $\tau$ only using the $EE$ QCS spectra. 

\subsubsection{Polarisation-only pixel-based likelihood} \label{sec:Likelihood_pixlike}
Under the assumptions that signal and noise are Gaussian and the noise covariance matrix is known accurately, it is possible to develop a pixel-based likelihood for low resolution CMB maps, see e.g. \citet{Page:2007}. This likelihood is formally exact, even for incomplete sky coverage, and is conceptually straightforward. A drawback of this likelihood is that the NCM has to be accurately known. Further, when using a pixel-based likelihood, it is difficult to assess whether or not a given result on $\tau$ is being driven by outliers at one or two multipoles.

The pixel-based polarisation-only likelihood is given by the probability of the data given the theoretical noise and signal model:
\begin{equation} \label{eq:pix_like}
    \mathcal{L}(\mat{d}) \equiv \mathcal{P}(\mat{d}|\mat{C}) = \frac{1}{\sqrt{|2\pi \mat{C}|}}\exp\left( -\frac{1}{2}(\mat{d}-\overline{\mat{d}})^\top \mat{C}\inv (\mat{d}-\overline{\mat{d}})\right) \ ,
\end{equation}
where $\mat{d}$ is the data vector and $\overline{\mat{d}}$ its smoothed mean, both containing the polarisation $Q$ and $U$ maps. This procedure can be trivially extended to account for temperature as well. However, in the present context we only focus on polarisation data. $\mat{C}$ is the covariance matrix which consists of a signal $\mat{S}$ and noise $\mat{N}$ component $\mat{C} = \mat{S} + \mat{N}$. The signal matrix can be constructed following the procedure outlined in e.g. \cite{Tegmark2001}. 

\section{Data} \label{sec:data}
We use various map products based on the time-ordered \Planck 2018 Low and High Frequency Instrument (LFI \& HFI) data\footnote{Recently, two new sets of \Planck maps have been introduced for LFI and HFI jointly \citep[\texttt{NPIPE},][] {2020arXiv200704997P} and for LFI only \citep[\texttt{BeyondPlanck},][]{BeyondPlanck:2020}. With a computation of corresponding noise covariance matrices, an analysis such as that presented here could be repeated with these maps.}. The data is processed with the following map-making algorithms: \madam for LFI \citep{Planck_LFI:2016, LFI_data:2018}, and \SRone and \SRtwo for HFI \citep{HFI_data:2018, 2019A&A...629A..38D}. The updated \SRtwo map-making algorithm reduces systematics in the polarisation data arising from analogue to digital converter non-linearities (ADCNL) \citep[][hereafter \citetalias{Aghanim:2018eyx}]{Aghanim:2018eyx}.

\subsection{Map compression}\label{sec:CMB_data}
\begin{figure}
    \centering
     \includegraphics[width=\columnwidth]{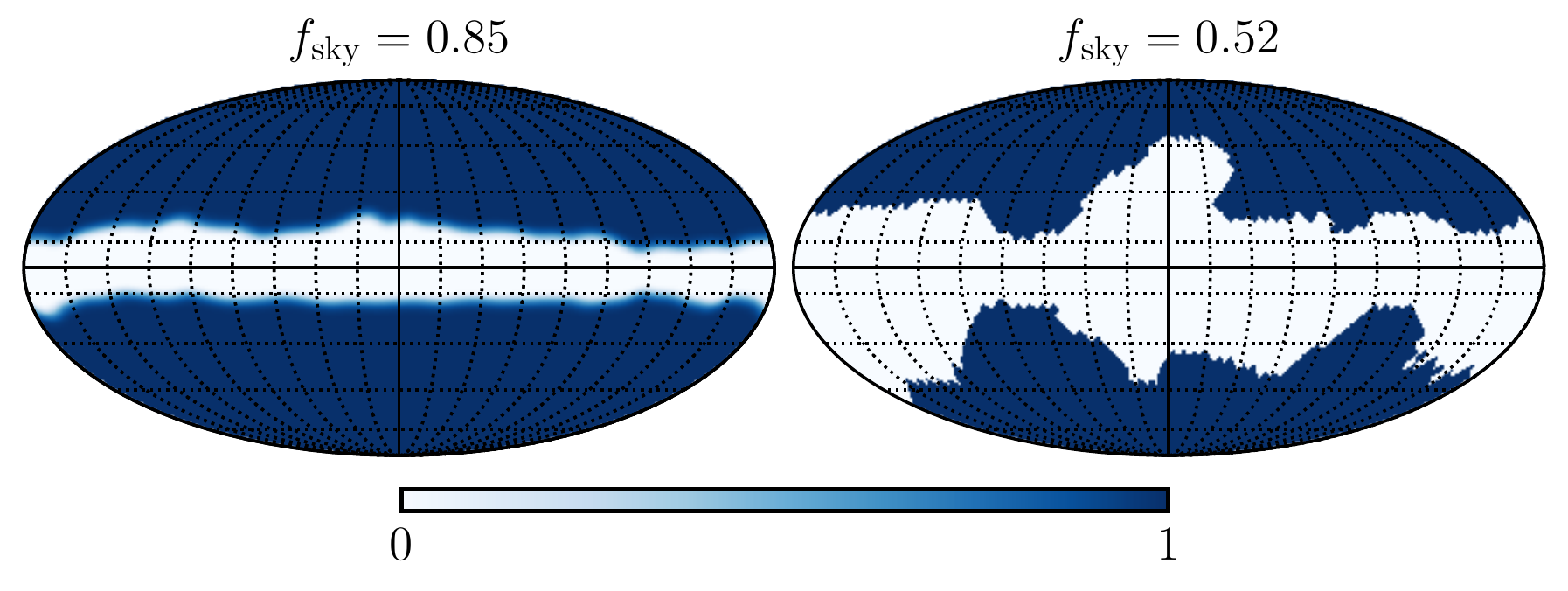}
    \vspace{-0.1in}
    \caption[Masks.]{Masks used in this paper. The quantity $f_{\rm sky}$ denotes the fraction of masked pixels, $f_{\rm sky} =\sum_{i=1}^{N_{\rm pix}} x_i /N_{\rm pix}^{\rm tot}$. The apodised mask ($0\leq x_i \leq 1$) with $f_{\rm sky}=0.85$ is used as a `processing' mask to perform the degrading of the high resolution maps to low resolution (as discussed in the text). The mask with $f_{\rm sky}=0.52$ is binary and used for foreground cleaning and power spectrum computations.}
    \label{fig:masks}
    \vspace{-0.1in}
\end{figure}

For the degradation process, we first apply an apodised mask with $f_{\rm sky}=0.85$ (as plotted in Fig.~\ref{fig:masks}) to the high resolution \Planck $Q$ and $U$ maps to suppress the Galactic plane region and  avoid smearing high amplitude foreground emission to high Galactic latitudes. The maps are then smoothed using the harmonic-space smoothing operator:
\begin{equation} \label{eq:harmonic_space_operator}
w(\ell) = \left\{ \begin{array}{ll}
 1 &,\ \ell \leq \ell_1\\
 \frac{1}{2}\left[ 1 + \cos{\pi\frac{\ell - \ell_1}{\ell_2-\ell_1}} \right] &,\ \ell_1 < \ell \leq \ell_2 \ ,\\
 0 &,\ \ell > \ell_2
\end{array} \right. \end{equation}
with $\ell_1 = N^{\rm lr}_{\rm side}$ and $\ell_2 = 3N^{\rm lr}_{\rm side}$. To be compatible with certain covariance matrix products, discussed below, we further smooth the maps with a $\texttt{HEALPix}$ pixel window function. Finally, we repixelise the maps at $N^{\rm lr}_{\rm side}= 16$ (with $N_{\rm pix}=3072$ pixels).

\subsection{Pixel-pixel noise covariance matrices}\label{sec:NCM}
The fidelity of our component separation procedure depends on how well the noise covariance matrices (NCMs) model the noise and systematics in the data. We closely follow the procedure used in \citetalias{Belsunce21}, extending the treatment from 100 and 143 GHz to all the LFI and HFI frequencies used in this paper. We  summarise only the key steps
here.

For the NCMs, we first fit a parametric model to the covariance matrix computed from $n_s$ end-to-end simulations for LFI and HFI data\footnote{We omit 100 end-to-end simulations, to test the robustness of our NCMs.}. The empirical noise and systematics only covariance matrix is constructed from simulations via $\mathbf{\Hat{N}} = n_s^{-1}\sum_i (\mathbf{n}_i - \mathbf{\Bar{n}})(\mathbf{n}_i - \mathbf{\Bar{n}})^{\top} $ where $\mathbf{n}_i$ are the simulated sky maps and $\mathbf{\Bar{n}}$ a smoothed template for $Q$ and $U$ Stokes parameters. The smoothed template corresponds to a smoothed mean of the maps (i.e.~a maximum likelihood solution of the map-making equation, see e.g.~\cite{Tegmark:1997}). 

We approach the problem of fitting a model to $\mathbf{\Hat{N}}$ as a maximum likelihood inference problem. We assume a Gaussian probability distribution for each noise realisation
\begin{align} \label{eq:likelihood_ncm}
    \mathcal{L} \equiv \mathbf{\mathcal{P}}(\mathbf{\hat{N}} \vert \mathcal{\mathbf{M}}) &= \prod_{i=1}^{n_s} \frac{1}{\sqrt{|2\pi \mathbf{M}|}}e^{-\frac{1}{2}(\mathbf{n}_i - \mathbf{\Bar{n}}) \mathbf{M}\inv(\mathbf{n}_i-\mathbf{\Bar{n}})^{\top}}  \ ,  
\end{align}
where  $\mat M$  is the model for the  noise covariance matrix. We assume that $\mat M$ consists of two terms
\begin{equation} \label{eq:noise_fit}
    \mat M  =  \alpha \mat N_0 + \mat{Y\mathbf{\Psi}Y}^\top,
\end{equation}
where the parameter $\alpha$ scales the low resolution (lr) map-making covariance matrix $\mat N_{0}$ and $\mat{Y \mathbf{\Psi}Y}^\top$ models additional large-scale effects. $\mat N_{0}$ is constructed from FFP8 end-to-end simulations \citep{2016A&A...594A...8P, 2016A&A...594A..12P} to capture the scanning strategy, detector white noise and `1/f'-type noise. The term $\mat{Y \mathbf{\Psi}Y}^\top$ models large-scale modes with parameters $\mathbf{\Psi}$ as described in more detail in \citetalias{Belsunce21}. To obtain a NCM, we minimise the negative log-likelihood of Eq.\ \eqref{eq:likelihood_ncm} with respect to $\alpha$ and $\mat M$.

\section{Results} \label{sec:results}
In this section, we present results of our component separation method on low-resolution polarised \Planck 2018 LFI and HFI data. In Sec.~\ref{sec:foreground_res} we discuss the recovered spectral index maps for electron synchrotron and thermal dust. We compare in Sec.~\ref{sec:tau_sec}  the effect of the presented foreground scheme to a spatially-uniform template cleaning method (TC) on cosmological parameter estimation. Therefore, we measure the optical depth to reionization, $\tau$, with a cross-spectrum based likelihood-approximation scheme (\momento) and a pixel-based likelihood (\pixlike), using our recovered CMB products. 

\subsection{Synchrotron and dust spectral index maps} \label{sec:foreground_res}

\begin{table}
\centering
{\def\arraystretch{1.2}\tabcolsep=5pt
\begin{tabular}{cccc}
\hline \hline
data     & case    &$\overline{\beta}_{\rm sync}$      & $\overline{\beta}_{\rm dust}$  \\ \hline
\SRone       & corrNCM + corrPrior & $-2.879 \pm 0.687$& $1.367 \pm 0.311$\\
             & corrNCM + diagPrior & $-2.931 \pm 0.590$& $1.398 \pm 0.272$\\
             & diagNCM + diagPrior & $-3.064 \pm 0.899$& $1.412 \pm 0.492$\\ 
\SRtwo       & corrNCM + corrPrior & $-2.833 \pm 0.620$& $1.429 \pm 0.236$\\ 
             & corrNCM + diagPrior & $-2.902 \pm 0.539$& $1.450 \pm 0.209$\\
             & diagNCM + diagPrior & $-2.980 \pm 0.854$& $1.460 \pm 0.402$\\
             \hline
\end{tabular}}
\caption{Summary of mean and standard deviations of the spectral indices for \SRone and \SRtwo maps using a mask with $f_{\rm sky}\simeq 0.52$. We show the three different cases that differ from each other by their degree of inter-pixel correlations in the noise covariance and prior matrices.}
\label{tab:beta_distribution}
\vspace{-0.1in}
\end{table}

In Table \ref{tab:beta_distribution} we present the mean $\overline{\beta}$ and standard deviation $\sigma(\beta)$ of the recovered spectral index maps. These are obtained for synchrotron and dust from our component separation method when using either \SRone or \SRtwo maps for HFI. Since the same \Planck 2018 LFI data is used in all cases, we expect differences in $\bdust$ and only minor changes in $\bsync$. 
We present our results for three cases with varying degrees of inter pixel correlations: (i) diagonal NCM and diagonal prior matrix (diagNCM + diagPrior); (ii) inter pixel correlations in the NCM and a diagonal prior matrix (corrNCM + diagPrior); (iii) inter pixel correlations in both the NCM and prior matrix (corrNCM + corrPrior). In our analyses, we impose wide priors of $\bsync=-3.0$ with $\sigma(\bsync)=2.0$ and $\bdust=1.5$ with $\sigma(\bdust)=1.0$. All the recovered spectral indices given in Table \ref{tab:beta_distribution} are in good agreement and return average values of $\overline{\beta}_{\rm sync} \approx -2.93$ and $\overline{\beta}_{\rm dust} \approx 1.42$ with large average spreads of $0.69$ and $0.32$ respectively. We adopt the results of the corrNCM + corrPrior case applied to full-mission HFI \SRtwo maps as our fiducial
analysis. For this case, we find the following means and standard deviations for the spectral indices: 
\begin{subequations}
\begin{align} 
\overline{\beta}_{\rm sync} &= -2.833\pm 0.620\ , \label{eq:bsync} \\
\overline{\beta}_{\rm dust} &= \phantom{-}1.429 \pm  0.236\ .  \label{eq:bdust}
\end{align}
\end{subequations}

\begin{figure}
    \centering
     \includegraphics[width=\columnwidth]{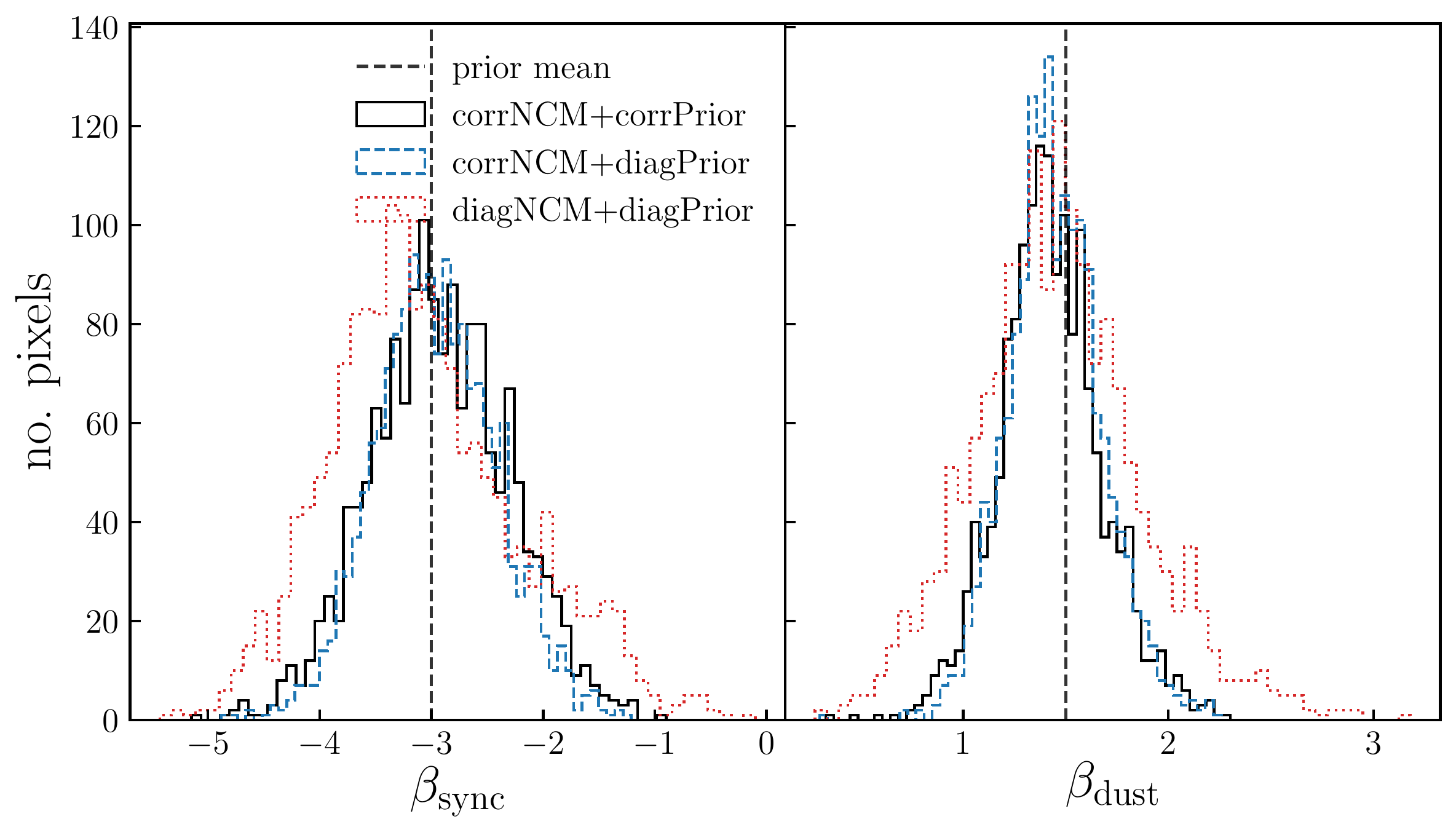}
    \vspace{-0.1in}
    \caption[]{Histograms of the spectral index maps shown in Fig.~\ref{fig:map_beta}. The input prior is shown as the dashed vertical black line for both spectral indices. The histograms of the three cases of the inter-pixel correlation we consider  are shown in solid black (corrNCM + corrPrior), dashed blue (corrNCM + diagPrior) and dotted red (diagNCM + diagPrior) lines.}
    \label{fig:beta_hist}
    \vspace{-0.1in}
\end{figure}

We show the histograms of the recovered synchrotron and dust spectral index maps in Fig.~\ref{fig:beta_hist} for the three cases when using HFI \SRtwo data. (The results for the same analyses using \SRone for HFI are shown in appendix \ref{sec:appendix_sroll1}.) As expected,
assuming independent pixels  (diagNCM + diagPrior) yields the largest dispersion  in $\beta$ values. Introducing inter-pixel correlations, we find
statistically insignificant  shifts in the  mean spectral indices, though the dispersions are reduced.  The effects of additionally including inter pixel-correlations in the prior matrix are minor compared to the case where these are only present in the NCM. 

Our mean recovered spectral indices given in Eqs. \eqref{eq:bsync}-\eqref{eq:bdust} are in agreement with, yet lower by $\sim 0.8 \sigma$ than, $\beta_{\rm dust}\approx 1.60$ obtained for LFI \citep{Svalheim:2020} and $\beta_{\rm dust}\approx 1.55$ for HFI data \citep{Planck_low-frequency_foregrounds:2016, Planck2020_compsep}. Similarly, the \SRtwo collaboration finds a narrow distribution for $\bdust$ \citep{Delouis:2021}. However, the spread of our spectral index maps is wider by a factor of $\sim 10-15$ for synchrotron and $\sim 4$ for dust. These differences stem mostly from our large prior widths that allow the $\beta$'s to explore a wider parameter space. Our priors are significantly broader than those adopted in previous studies, see e.g.~\citet{Dunkley:2009fg, Armitage-Caplan2011,Fuskeland:2014,Svalheim:2020}, which are in alignment with the \Planck priors \citep{Planck_diffuse_LFI:2016, Planck2020_compsep}\footnote{Our broad spreads in the spectral indices are consistent, however, with the `prior-free' tests performed in the latter \Planck paper for which one or the other of the tight Gaussian priors is removed.}. These analyses use tight Gaussian priors of $\bsync=-3.10\pm 0.10$ and $\bdust=1.56\pm 0.10$. We test our component separation method for these priors and recover similar prior-driven results for \SRtwo full-mission data $\Bar{\beta}_{\rm sync} =-3.000\pm 0.008$ and $\Bar{\beta}_{\rm dust}=1.494\pm0.028$ using the pixel-by-pixel method. The recovered spectral index maps are almost identical when including inter pixel correlations. Such priors are so tight that they do no allow any significant exploration of the parameter space, which results in almost automatically rejecting the hypothesis of spectral index variations across the sky. Therefore we emphasise that \emph{previous results are strongly prior dominated}. We next investigate angular structure in our recovered spectral index maps.

\begin{figure*}
    \centering
     \includegraphics[width=\textwidth]{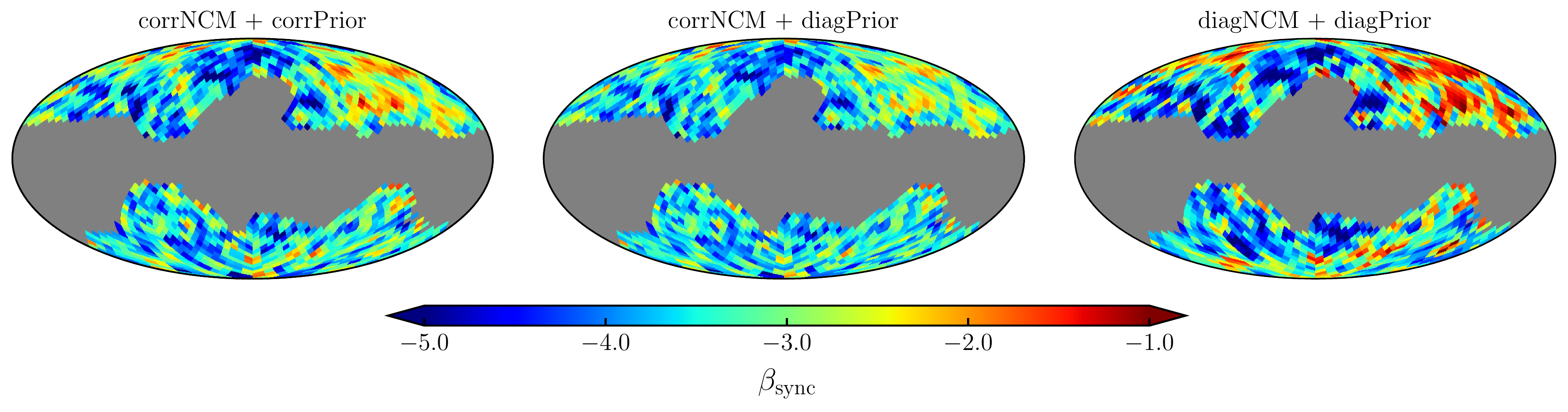}
     \includegraphics[width=\textwidth]{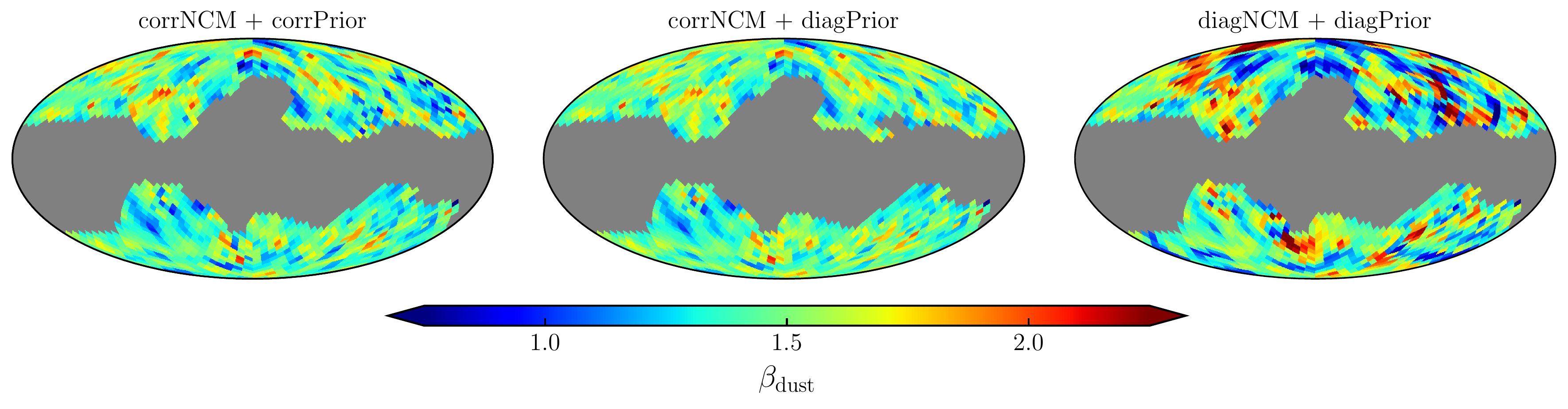}
    \vspace{-0.1in}
    \caption[]{Distribution of synchrotron (upper panel) and dust (lower panel) spectral indices at the map level for full-mission \SRtwo maps. The left column shows the fully correlated `corrNCM + corrPrior' case, the middle column shows the `corrNCM + diagPrior' case, and the right one shows the pixel-by-pixel `diagNCM + diagPrior' case. Masked pixels are shown in grey. The corresponding histograms of the spectral index maps are shown in Fig.~\ref{fig:beta_hist} and the associated means and standard deviations are given in Table \ref{tab:beta_distribution}.}
    \label{fig:map_beta}
    \vspace{-0.1in}
\end{figure*}

\begin{figure*}
    \centering
     \includegraphics[width=\textwidth]{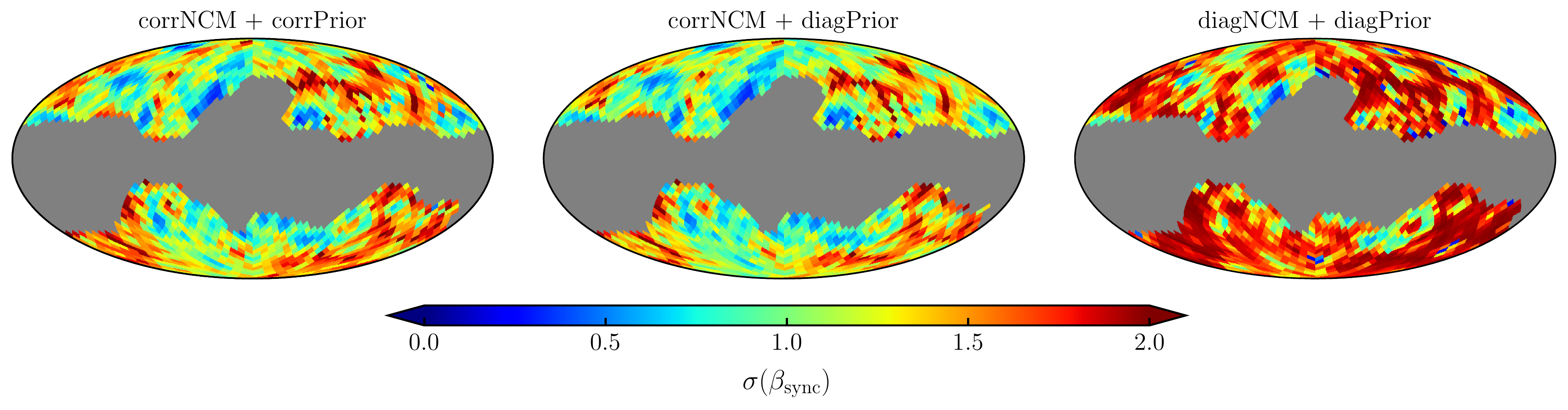}
     \includegraphics[width=\textwidth]{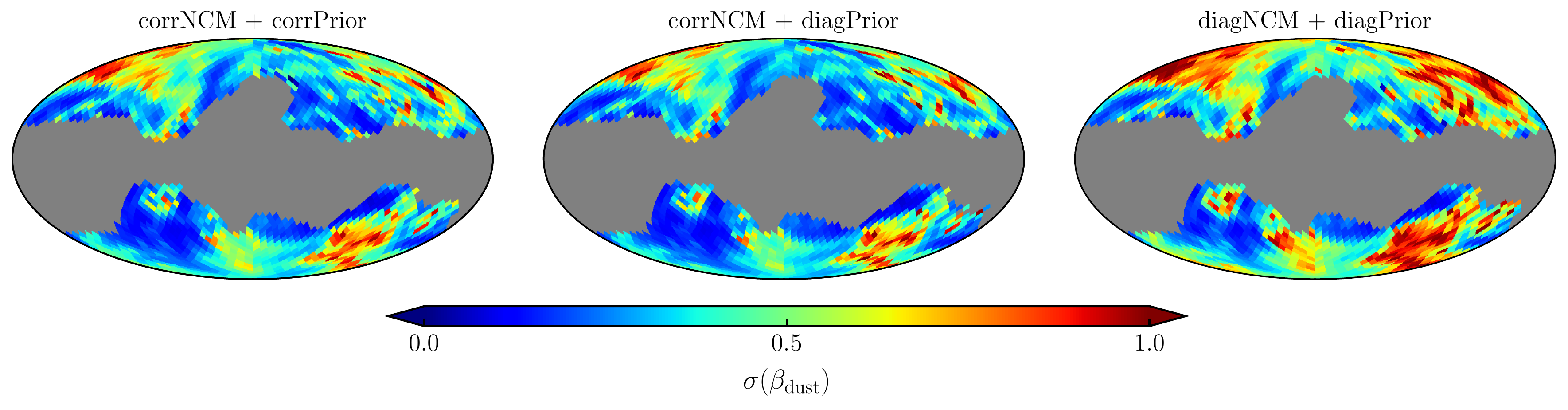}
    \vspace{-0.1in}
    \caption[]{Maps of uncertainties for synchrotron (upper panel) and dust (lower panel) spectral indices for full-mission \SRtwo maps. These are obtained from the output of the Bayesian component separation method, as the inverse of the curvature matrix after the scheme has converged. The ordering of the maps is analogous to Fig.~\ref{fig:map_beta}. Masked pixels are again shown in grey.}
    \label{fig:map_beta_noise}
    \vspace{-0.1in}
\end{figure*}

\subsection{Spectral index variations}
In Fig.~\ref{fig:map_beta} we show the recovered synchrotron and dust spectral index maps. One immediately notices the apparent feature in the upper right quadrant with particularly noticeable outliers in the spectral indices in the independent pixel case\footnote{This sky region is where differences in the polarization maps between different processing pipelines are particularly noticeable at 30 GHz; see figs.\ 42 and 50 of \citet{2020arXiv200704997P}. Indeed, the time-dependent gain calibration at 30 GHz is difficult given the limited number of radiometers. This may lead to temperature-to-polarisation leakage in regions of the sky where the temperature fluctuations are largest, as set by the CMB dipole. Additionally, the upper right quadrant of the sky map coincides partially with effects stemming from analogue-to-digital converter non-linearities \citepalias[ADCNL;][]{Aghanim:2018eyx} in the HFI data. Thus it would be interesting to see how our recovered spectral index maps might change when using alternative data products as discussed in Sec.\ \ref{sec:data}.}. These pixels are located in the tails of their respective histograms. To investigate further, we look at the uncertainties our method yields for all of the fitted parameters with their covariance being given by the inverse of the curvature matrix evaluated at the solution. In Fig.~\ref{fig:map_beta_noise} we plot the square root of the diagonals of the spectral index covariance matrices. We note the increased uncertainties in the recovered spectral indices in the upper right quadrant of the sky map, decreasing their apparent significance. In fact, in the diagNCM + diagPrior case one can see that the recovered uncertainties are approaching their maximal values (that of the prior) over much of the sky for the synchrotron spectral index; this indicates that the data is not constraining $\bsync$ at all in the upper right quadrant. 

\begin{figure}
    \centering
     \includegraphics[width=0.9\columnwidth]{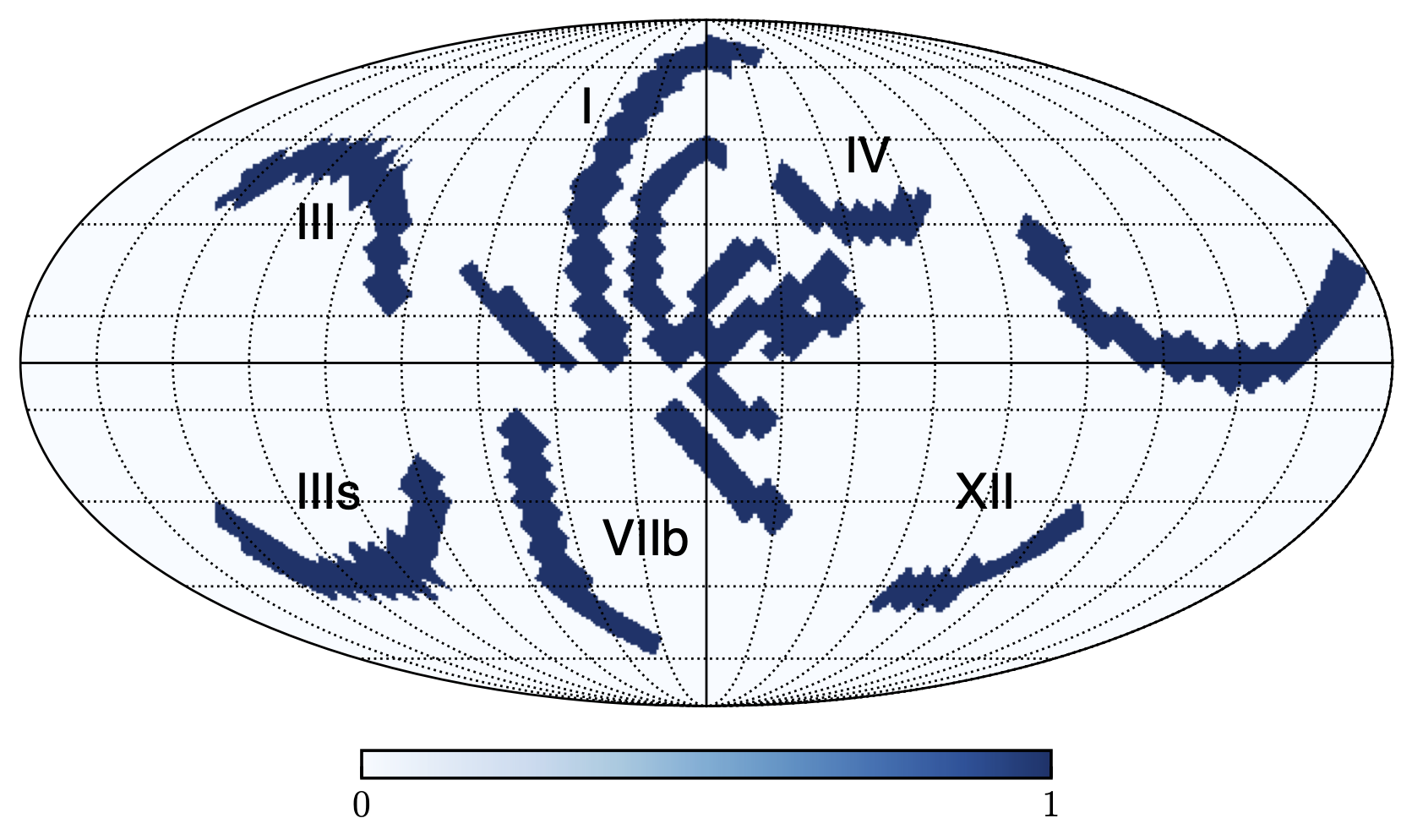}
    \vspace{-0.1in}
    \caption[Masks.]{Map showing the radio loops in the 480 MHz map shown in \citet{Haslam1982} and adapted from Fig. 2 in \citet{Vidal2015}, downgraded to $N_{\rm side}=16$. The loops follow the maximum brightness points along each filament of the 480 MHz high-pass filtered map. The six named loops (I, III, IIIs, IV, VIIb, XII), see e.g.~\citet{Dickinson_2018} for a review, are investigated in looking for spatial variations of the spectral indices.}
    \label{fig:mask_loop}
    \vspace{-0.1in}
\end{figure}

More physically, the synchrotron spectral index maps in Fig.~\ref{fig:map_beta} display ring-like structures with steeper $\bsync$. These features correlate with the positions of radio loops, shown in Fig.~\ref{fig:mask_loop}, which are nearby structures within a few hundred parsecs (for a review, see e.g.~\citet{Dickinson_2018}). Electron synchrotron radiation in these loops comes from young, fast cosmic rays which are remnants of supernovae.  The uncertainties from the curvature matrix, shown in Fig.~\ref{fig:map_beta_noise}, are smaller within the loops, effectively increasing the statistical significance of spectral index variations within the loops. 

\begin{table*}
\centering
{\def\arraystretch{1.2}\tabcolsep=10pt
\begin{tabular}{@{}ll*{4}{c}@{}}\hline\hline
method& data &\multicolumn{2}{c}{\SRtwo}&\multicolumn{2}{c}{\SRone}\\\hline
& & $\overline{\beta}_{\rm sync}$& $\overline{\beta}_{\rm dust}$ & $\overline{\beta}_{\rm sync}$& $\overline{\beta}_{\rm dust}$  \\ \cmidrule(lr){3-4} \cmidrule(lr){5-6}
corrNCM + corrPrior & I &$-3.228 \pm 0.496$& $1.447 \pm 0.291$& $-3.264 \pm 0.630$&$1.397 \pm 0.377$\\
                    & III &$-3.236 \pm 0.674$& $1.388 \pm 0.243$& $-3.589 \pm 0.697$&$1.451 \pm 0.241$\\
                    & IIIs  &$-3.001 \pm 0.449$& $1.356 \pm 0.193$  &$-3.150 \pm 0.504$&$1.442 \pm 0.180$\\
                    & VIIb  &$-3.043 \pm 0.519$& $1.408 \pm 0.176$  &$-3.025 \pm 0.555$&$1.339 \pm 0.180$\\
                    & XII  &$-2.996 \pm 0.454$& $1.095 \pm 0.269$  &$-2.774 \pm 0.521$&$1.485 \pm 0.217$\\
                    & IV  &$-2.950 \pm 0.692$& $1.293 \pm 0.249$  &$-2.966 \pm 0.780$&$1.256 \pm 0.273$\\
                    \rule{0pt}{3ex} 
corrNCM + diagPrior & I &$-3.319 \pm 0.423$& $1.457 \pm 0.253$ &$-3.208 \pm 0.523$&$1.411 \pm 0.333$\\
                    & III &$-3.328 \pm 0.574$& $1.539 \pm 0.203$ &$-3.476 \pm 0.600$&$1.453 \pm 0.220$\\
                    & IIIs  &$-3.054 \pm 0.419$& $1.372 \pm 0.192$  &$-3.118 \pm 0.457$&$1.444 \pm 0.177$\\ 
                    & VIIb  &$-3.108 \pm 0.451$& $1.417 \pm 0.164$  &$-3.035 \pm 0.499$&$1.370 \pm 0.164$\\ 
                    & XII  &$-3.096 \pm 0.377$& $1.493 \pm 0.187$  &$-2.821 \pm 0.482$&$1.495 \pm 0.197$\\ 
                    & IV  &$-3.063 \pm 0.634$& $1.330 \pm 0.225$  &$-2.977 \pm 0.696$&$1.289 \pm 0.256$\\
                    \rule{0pt}{3ex} 
diagNCM + diagPrior & I &$-3.606 \pm 0.712$& $1.339 \pm 0.247$&$-3.772 \pm 0.816$&$1.611 \pm 0.526$\\
                    & III &$-3.720 \pm 0.732$& $1.645 \pm 0.315$ &$-3.881 \pm 0.705$&$1.417 \pm 0.324$\\
                    & IIIs  &$-3.418 \pm 0.393$& $1.471 \pm 0.242$  &$-3.718 \pm 0.493$&$1.656 \pm 0.282$\\ 
                    & VIIb  &$-3.578 \pm 0.492$& $1.545 \pm 0326$  &$-3.481 \pm 0.664$&$1.476 \pm 0.375$\\ 
                    & XII  &$-3.006 \pm 0.580$& $1.533 \pm 0.394$  &$-2.787 \pm 0.593$&$1.469 \pm 0.345$\\ 
                    & IV  &$-3.562 \pm 0.728$& $1.121 \pm 0.197$  &$-3.604 \pm 0.855$&$1.373 \pm 0.273$\\ \hline

\end{tabular}}
\caption{Summary of spectral index distributions for synchrotron and dust in six relevant radio loops, shown in Fig.~\ref{fig:mask_loop}, for full-mission \SRone and \SRtwo data. We compare three different cases: (i) corrNCM + corrPrior, (ii) corrNCM + diagPrior, and (iii) diagNCM + diagPrior.  The spectral index maps are shown for \SRtwo in Fig.~\ref{fig:map_beta} and for \SRone in Fig.~\ref{fig:map_beta_sroll1}.}
\label{tab:beta_distribution_loops}
\vspace{-0.1in}
\end{table*}

To quantify these features further, we quote the mean and standard deviations of the spectral index maps within the radio loops in Table \ref{tab:beta_distribution_loops}. For synchrotron, the means of loops I, III, IIIs, VIIb, and IV deviate from the mean across the unmasked sky by just under one standard deviation $\sigma$ {\it for a single pixel}. With each loop containing $n\sim 30-50$ pixels and if each pixel's spectral index is assumed to be independent, one would expect the mean for each loop to agree with that across the whole sky up to a standard deviation $\sigma/\sqrt{n}$. Thus, we assess the evidence for spatial variations to be at the two- to four-sigma level. As expected, the pixel-by-pixel procedure shows stronger spatial variations of $\bsync$ than the correlated methods. This comes from the inter pixel correlations effectively smoothing out features and penalising neighbouring pixels with (strongly) dissimilar $\beta$-values. In contrast to synchrotron, visual inspection of the dust spectral index maps, shown in Fig.~\ref{fig:map_beta}, and the quantitative test in the loop regions, quoted in Table \ref{tab:beta_distribution_loops}, reveal no clear evidence for dust spectral index variations across the sky.

Spectral index variations have been measured in e.g.~\citet{Fuskeland:2014} who with tight priors find evidence for weak spatial variations in polarised synchrotron emission using the nine-year \emph{Wilkinson Microwave Anisotropy Probe} data \citep[WMAP,][]{Bennett:2003}.  Their analysis, similar to the Gibbs sampling technique in \citet{Eriksen2008} and \citet{Dunkley:2009}, finds an increasingly steep spectral index for higher Galactic latitudes, ranging from $\beta_{\rm plane} = -2.98\pm0.01$ to $\beta_{\rm high-lat}=-3.12\pm0.04$. Recent results on LFI \Planck data by \citet{Svalheim:2020} report similar findings.  In appendix \ref{app:lat_dependence} we test for latitudinal spectral index variations but do not see a significant effect for either $\bsync$ or $\bdust$. 

\subsection{Effect on cosmological parameter estimation} \label{sec:tau_sec}
An important aim of the present investigation is to quantify the effect of spatially varying Galactic polarised foregrounds on cosmological parameter investigation. Therefore, we measure the optical depth to reionization parameter, $\tau$, comparing our findings here using our Newton-Raphson component separation procedure (henceforth denoted by NR) to \citetalias{Belsunce21} which used simple  template cleaning (henceforth denoted by TC). The TC method cleaned the \SRtwo signal channels at 100 and 143 GHz with uniformly scaled versions of 30 GHz (for synchrotron) and 353 GHz (for dust) half-mission maps using the coefficients of table 1 in \citetalias{Belsunce21}. The NR analysis uses sets of either full-mission or half-mission maps.
To explore the cosmological parameter space, we perform one-dimensional parameter scans in $\tau$. Therefore, we generate theoretical power spectra with the base \LCDM cosmology of \citepalias{Aghanim:2018eyx}: $H_0 = 67.04$, $\Omega_{\rm b}h^2 = 0.0221$, $\Omega_{\rm c}h^2 = 0.12$, $\Omega_{\nu}h^2 = 0.00064$, $\theta_*=1.0411$, $n_{\rm s}=0.96$, and keeping $10^9 A_s e^{-2 \tau} =1.870$ fixed. A fiducial $\tau=0.060$ is assumed for forming the quadratic cross-spectra (QCS) and for the scans $\tau$ ranges from $0.01$ to $0.2$ with $\Delta \tau=0.001$.

\begin{figure}
    \centering
     \includegraphics[width=\columnwidth]{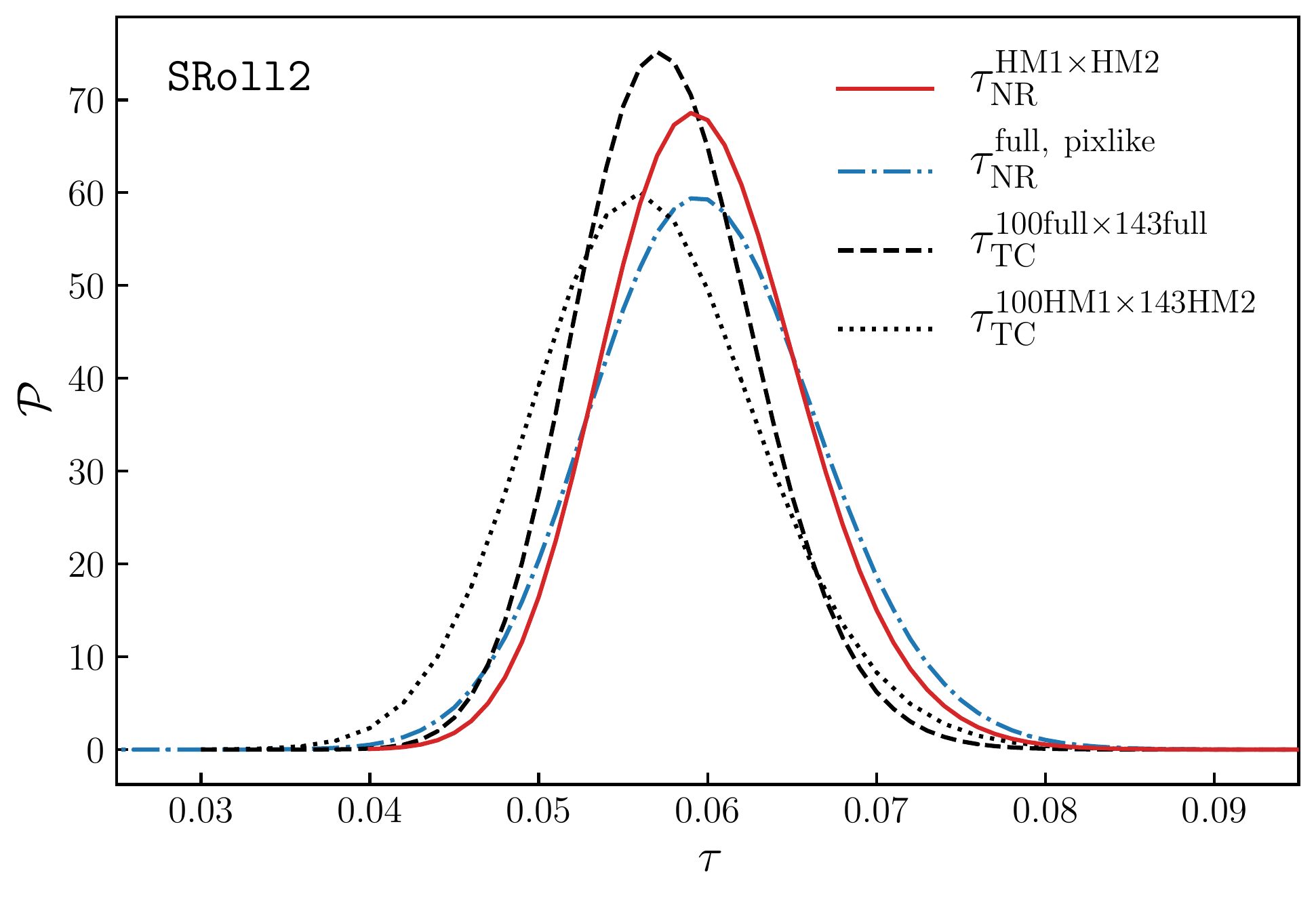}
    \vspace{-0.1in}
    \caption[tau]{Posteriors for the optical depth to reionization using \SRtwo data for HFI. We compare the CMB maps obtained from the presented Bayesian component separation method (labelled NR) and template-cleaned (labelled TC) maps. We perform a one-dimensional parameter scan over $\tau$ using a pixel-based likelihood (\pixlike) and a cross-spectrum-based likelihood (\momento).}
    \label{fig:tau}
    \vspace{-0.1in}
\end{figure}

In Fig.~\ref{fig:tau} we compare the posteriors for $\tau$ from NR-cleaned \SRtwo data using the two likelihoods introduced in Sec.~\ref{sec:likelihoods}. 
We produce the two independent CMB maps required for the likelihood-approximation scheme \momento, which uses $EE$ cross-spectra, by running the NR procedure twice, once on half-mission 1 (HM1) input maps and once on half-mission 2 (HM2) input maps. We label the obtained posterior $\mathrm{HM1}\times \mathrm{HM2}$. The pixel-based likelihood \pixlike uses $Q$ and $U$ full-mission maps to constrain $\tau$, giving a posterior we label `$\mathrm{full,\, pixlike}$'. We use the appropriate block of the inverse of the recovered curvature matrix as the effective noise covariance matrix for each map in the likelihoods, propagating foreground-cleaning uncertainties into the $\tau$ estimates. We also show two posteriors obtained using TC cleaning of \SRtwo from \citetalias{Belsunce21}. The $100\times 143$ GHz full-mission posterior was the baseline result, and the $100\,\mathrm{HM1}\times 143\,\mathrm{HM2}$ posterior relates to the NR-cleaned half-mission posterior ($\mathrm{HM1}\times \mathrm{HM2}$) in its use of half-mission data.

The means and standard deviations of the four posteriors are:
\begin{subequations}
\begin{align}
    \tau_{\rm NR} &= 0.0598 \pm 0.0059\,  \ (\momento\, \mathrm{HM1}\times\mathrm{HM2}),
    \label{eq:momento_NR_HM}\\
    \tau_{\rm NR} &= 0.0599 \pm 0.0066\,  \ (\pixlike\ \rm full), \label{eq:pixlike_NR_full}\\
    \tau_{\rm TC} &= 0.0581 \pm 0.0055\,  \ (\momento\, \mathrm{100\, full}\times\mathrm{143\, full}), \label{eq:momento_TC_full}\\
    \tau_{\rm TC} &= 0.0565 \pm 0.0068\,  \ (\momento\   \mathrm{100\, HM1}\times\mathrm{143\, HM2}). \label{eq:momento_TC_HM}
\end{align}
\end{subequations}
It is reassuring to see that the two likelihoods give consistent results when using NR cleaning, and that both of these analyses are consistent with the TC-based results presented in \citetalias{Belsunce21}. We stress again our use of wide priors in the NR method. Indeed, applying tight priors leads to results on $\tau$ that are higher than the TC-cleaned ones by several $\sigma$. Such results are expected as the freedom to reduce foreground residuals in the recovered CMB maps is constrained by a tight prior.

\section{Summary and Conclusion} \label{sec:conclusion}

Foregrounds pose a major challenge to cosmological parameter estimation from large angular scale ($2 \leq \ell \leq 30$) polarised CMB data. We have presented a Bayesian parametric foreground fitting framework to disentangle the primary CMB and the Galactic polarised foregrounds from one another. We modelled electron synchrotron radiation as a power law and thermal dust emission as a modified blackbody in frequency space. Our method allows for inter-pixel correlations in the noise covariance matrices of the input maps and for a physically motivated spatial correlation length in the spectral index prior matrices. We used six frequency channels from polarised \Planck 2018 LFI and HFI data and found evidence for a spatially varying spectral index for synchrotron but not for dust. The means and dispersions of both spectral indices are given in Eqs.~\eqref{eq:bsync} and \eqref{eq:bdust}. From an analysis of  our synchrotron spectral index maps, we find significantly steeper indices  in the locations of radio loops.

We have investigated  whether the optical depth to reionization $\tau$ measured from the \Planck 2018 LFI and \SRtwo HFI polarisation maps
is sensitive to the method of polarized foreground removal.
We have applied  the  Bayesian parametric foreground component separation method developed in this paper, using both a  cross-spectrum-based likelihood \momento and a pixel-based likelihood \pixlike, to derive the  $\tau$ posteriors given in Eqs.~\eqref{eq:momento_TC_full}-\eqref{eq:pixlike_NR_full}. 
  These results are in excellent agreement with the simple uniform template cleaning method adopted in B21.
  
We do not anticipate any further methodological advances that would lead to substantive improvements in \Planck polarization maps at  large angular scales. Next-generation space-borne CMB experiments, however, will measure the sky with higher signal-to-noise and their analysis will be sensitive to the spatial variations of polarised foreground properties. Foreground cleaning and inference methods such as those presented here  will thus be necessary for accurate cosmological inference from upcoming polarised CMB surveys with their ambitious science targets such as the measurement of primordial $B$-modes. This will be especially acute at lower CMB frequencies in which synchrotron is a significant contaminant which has, as we have seen, a complex angular  structure.

\section*{Acknowledgements}
The authors thank the \Planck Collaboration for their tremendous efforts in producing this data set and especially thank the Bware team for their \SRtwo products. Further the authors thank C.\ Dickinson and M.\ Vidal for providing the radio loop map used in the analysis, shown in Fig.~\ref{fig:mask_loop}, and thank J.\ M.\ Delouis, J.\ L.\ Puget, and R.\ Keskitalo for useful discussions. RdB acknowledges support from the Isaac Newton Studentship, Science and Technology Facilities Council (STFC) and Wolfson College, Cambridge. SG acknowledges the award of a Kavli Institute Fellowship at KICC. We acknowledge the use of: \texttt{HEALPix} \citep{2005ApJ...622..759G} and \texttt{CAMB} (\url{http://camb.info}). This work was performed using the Cambridge Service for Data Driven Discovery (CSD3), part of which is operated by the University of Cambridge Research Computing on behalf of the STFC DiRAC HPC Facility (\url{www.dirac.ac.uk}). The DiRAC component of CSD3 was funded by BEIS capital funding via STFC capital grants ST/P002307/1 and ST/R002452/1 and STFC operations grant ST/R00689X/1. DiRAC is part of the National e-Infrastructure.

\section*{Data availability}
The data underlying this article will be shared on reasonable request. The \Planck maps and end-to-end simulations are publicly available. The code is publicly available on GitHub: \url{https://github.com/roger2b/Bayesian-parametric-fitting}.



\bibliographystyle{mnras}
\bibliography{references} 


\appendix
\section{Spectral index models} \label{sec:appendix_spec_idx}
If radiation at a frequency $\nu$ has brightness $I_{\nu}$, one defines a brightness temperature $T_B$ from this as
\begin{equation}
    I_{\nu} = \frac{2 k_B \nu^2}{c^2} T_B(\nu)\ , 
\end{equation}
where $k_B$ is the Boltzmann constant, and $c$ the speed of light. For a blackbody
\begin{equation}
    I_{\nu} = \frac{2h\nu^3}{c^2}\frac{1}{e^x-1} \ ,
\end{equation}
with $x=h\nu / k_B T$, where $h$ is Planck's constant. Thus one can relate thermodynamic and brightness temperature variations through
\begin{equation}
    \frac{\Delta T_B}{\Delta T} = \frac{x^2 e^x}{\left(e^x-1\right)^2} \equiv g(x)\ .
\end{equation}

We model the  synchrotron radiation as a power law and so take $\Delta T_{\rm B}^{\rm sync} \propto \nu^{\beta_{\rm s}}$. We characterise the dust spectral energy distribution (SED) by a modified blackbody
\begin{equation}
    I_{\nu}^{\rm d} \propto \left(\frac{\nu}{\nu^{\rm d}_0} \right)^{\beta_{\rm d}} \frac{\nu^3}{\exp{(h\nu / k_{\rm B}T_{\rm d}) -1}}\ ,
\end{equation}
where we assume $T_{\rm d}=22.7$ K \citep{EG19} for the dust temperature. Therefore, the brightness temperature is
\begin{equation}
    \Delta T_B^{\rm dust} \propto \left(\frac{\nu}{\nu^{\rm d}_0} \right)^{\beta_{\rm d}+1} f(y)\ ,
\end{equation}
where we have defined $f(y)\equiv 1/(e^y-1)$ with $y=h\nu / k_B T_{\rm d}$. The numerical values are given in Table \ref{tab:conversion_fac}, where we adopt a fiducial frequency for dust of $\nu_0^{\rm d} = 100$ GHz. The full foreground model in thermodynamic temperature units is given in Eq.~\eqref{eq:model_fg}. 

\begin{table}
\centering
{\def\arraystretch{1.2}\tabcolsep=5pt
\begin{tabular}{lcccccccc}
\hline \hline
$\nu$     & 30 &44& 70 & 100 &143 &217 &353\\ 
$\nu_{\rm cen.}$ & 28.4 &44.1& 70.4 & - & - & - & -  \\
$\nu_{\rm eff.}$ & - & - & - &$101.3$ &$142.7$ &$221.9$ &$361.3$\\ 
$\lambda$     & 10 & 6.8 & 4.2 & 3.0 & 2.1 & 1.4 & 0.85\\ 
$\Delta T/\Delta T_B$     & 1.02 & 1.05 & 1.13 & 1.29 & 1.65 & 3.13 & 14.29\\
$f(y_0^{\rm d})/f(y)$     & 3.80 & 2.40 & 1.47 & 0.985 & 0.669 & 0.393 & 0.205\\\bottomrule
\end{tabular}}
\caption{Conversion factors between antenna and thermodynamic temperature for the nominal frequencies quoted in the first row. In addition to the nominal frequencies $\nu$, we also quote the central \citep{Planck2013_LFI} and effective \citep{Planck2013_HFI} frequencies for LFI and HFI. The frequencies $\nu$ are given in units of GHz and wavelengths $\lambda$ in mm.}
\label{tab:conversion_fac}
\vspace{-0.1in}
\end{table}

\section{Spectral index variations in \SRone data} \label{sec:appendix_sroll1}
\begin{figure*}
    \centering
     \includegraphics[width=\textwidth]{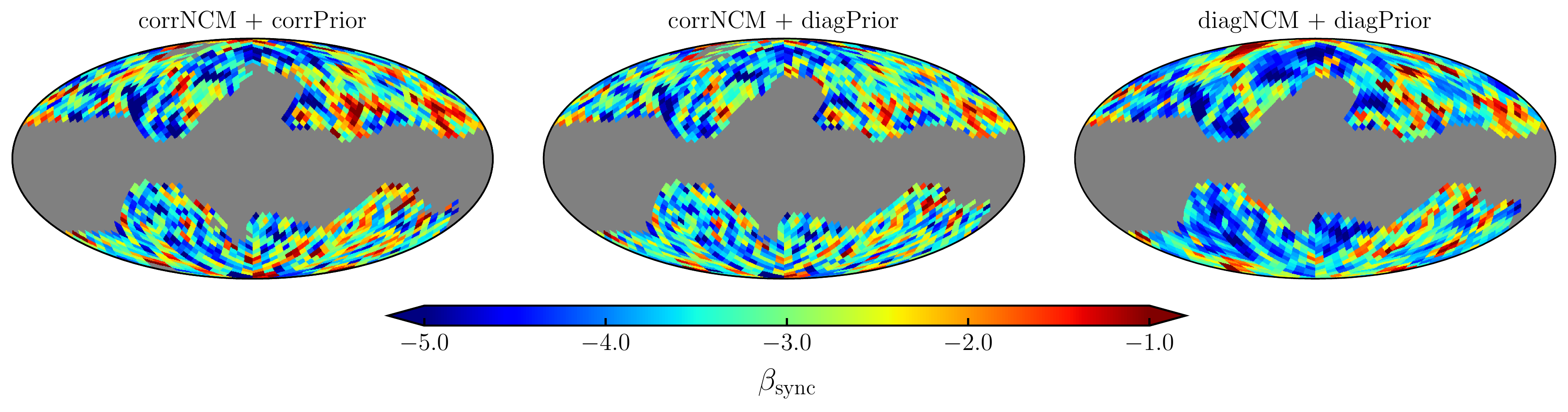}
     \includegraphics[width=\textwidth]{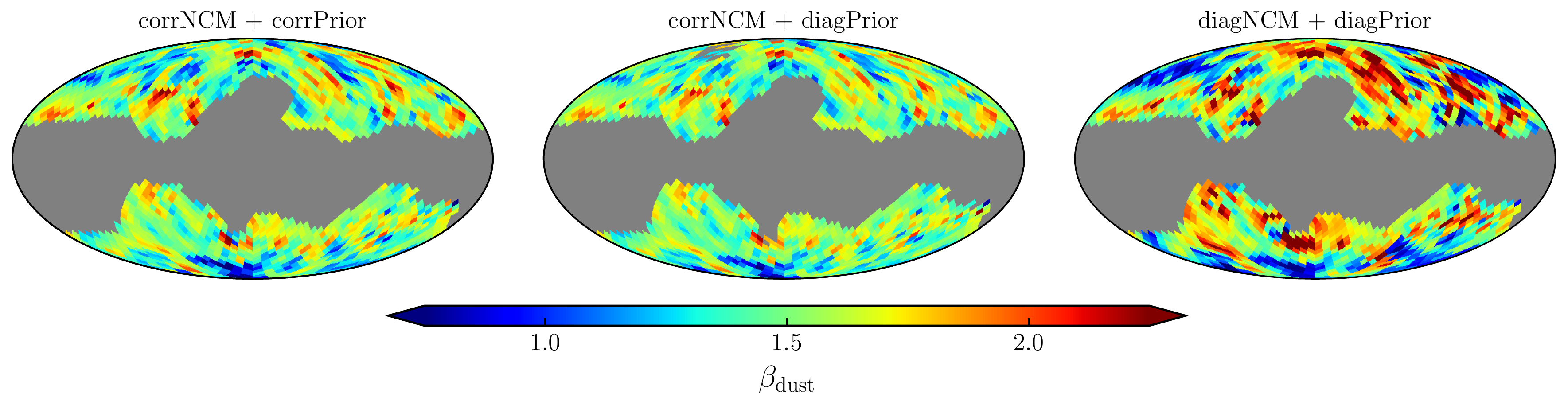}
    \vspace{-0.1in}
    \caption[]{Distribution of synchrotron (upper panel) and dust (lower panel) spectral index maps using \SRone for HFI for the three inter-pixel correlation cases considered. Masked pixels are shown in grey. The means and standard deviations of the maps are given in Table \ref{tab:beta_distribution}.}
    \label{fig:map_beta_sroll1}
    \vspace{-0.1in}
\end{figure*}
\begin{figure}
    \centering
     \includegraphics[width=\columnwidth]{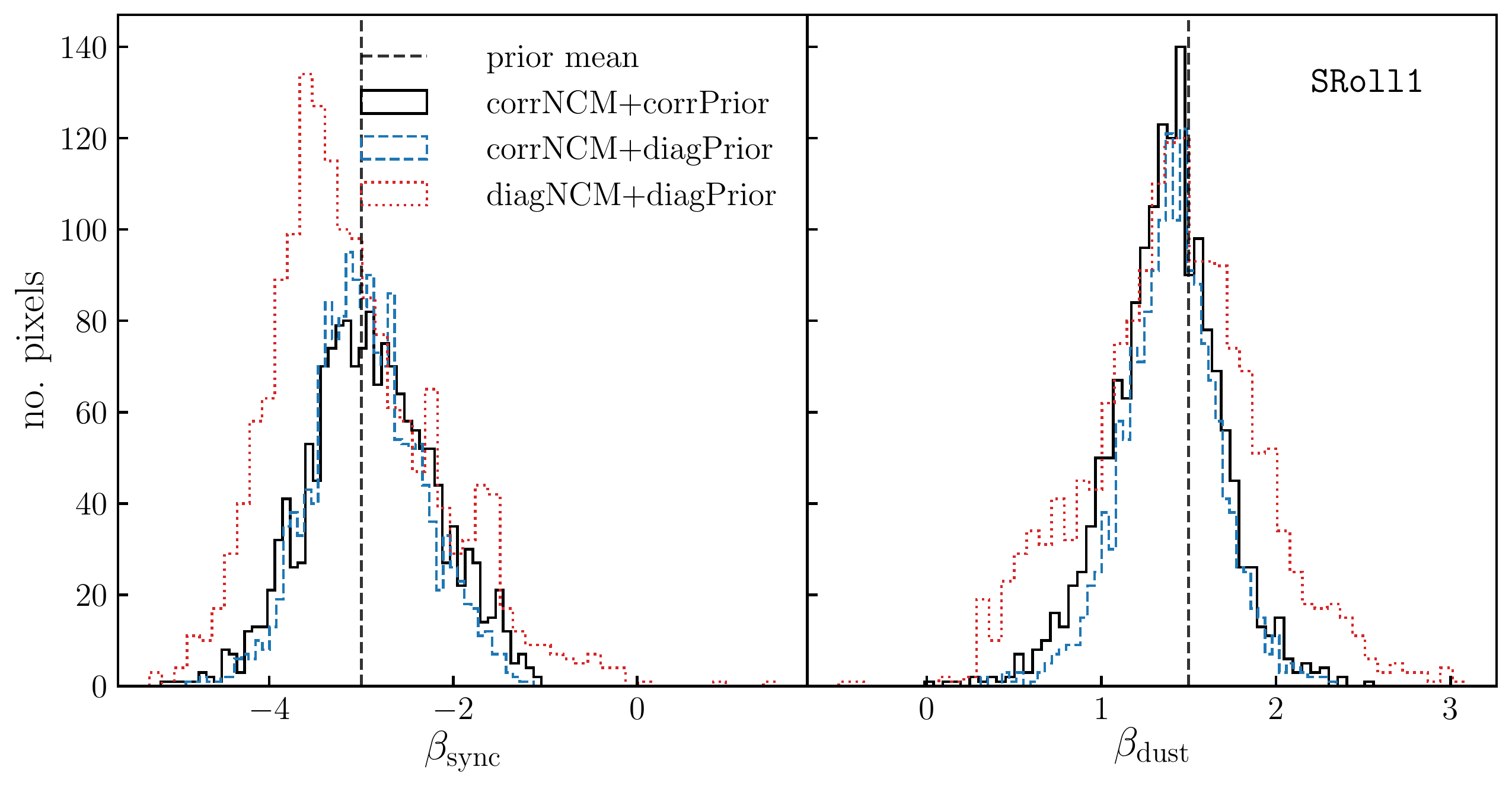}
    \vspace{-0.1in}
    \caption[]{Histogram of the spectral index maps shown in Fig.~\ref{fig:map_beta_sroll1}. The input priors are shown as dashed vertical black lines for both spectral indices. The histograms of the three cases are shown in solid black (correlated noise + correlated prior), dashed blue (correlated noise + diagonal prior) and dotted red (diagonal noise + diagonal prior). }
    \label{fig:beta_hist_sroll1}
    \vspace{-0.1in}
\end{figure}

In Fig.~\ref{fig:map_beta_sroll1}, we show the spectral index maps obtained in the three cases introduced in Sec.~\ref{sec:correlations} using \SRone data in place of \SRtwo data for HFI. The corresponding histograms are shwon in Fig.~\ref{fig:beta_hist_sroll1}. 

\section{Mask dependence of the \SRone results} \label{app:mask_dependence}
This appendix discusses the effect of increasing the fraction of unmasked sky pixel $f_{\rm sky}$ in the presented component separation method. For this test we used \SRone data for HFI. Inter pixel correlations cause high amplitude pixels closer to the Galactic plane to affect areas away from the Galactic plane, potentially leading to convergence issues. Ignoring these correlations in the noise and the prior matrices, we test using a mask with sky fraction $f_{\mathrm{sky}}\simeq 0.66$ (and a simple noise variance pattern suitable for this mask).

\begin{figure}
    \centering
     \includegraphics[width=\columnwidth]{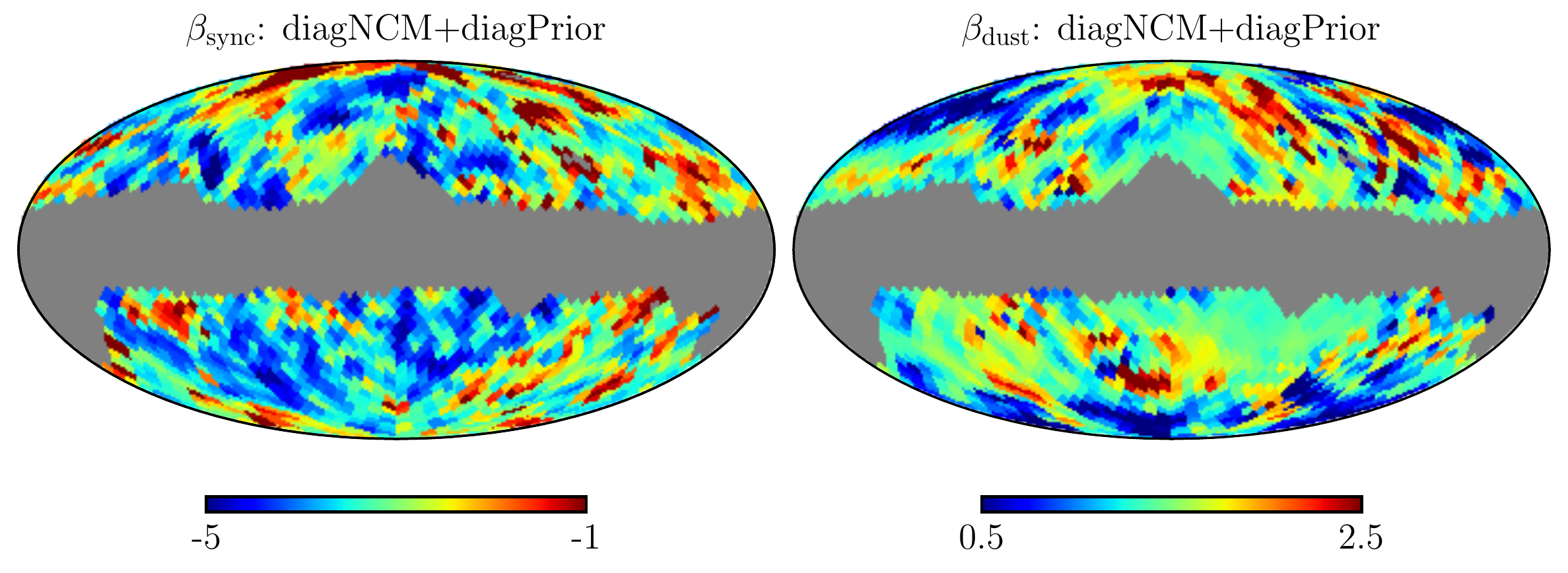}
    \vspace{-0.1in}
    \caption[]{Distribution of synchrotron (left) and thermal dust (right) spectral indices across the sky using \SRtwo data for HFI. The masked pixels are shown in grey with a fraction of $f_{\rm sky}=0.66$ unmasked pixels.}
    \label{fig:pix_mask_beta}
    \vspace{-0.1in}
\end{figure}

\begin{figure}
    \centering
     \includegraphics[width=\columnwidth]{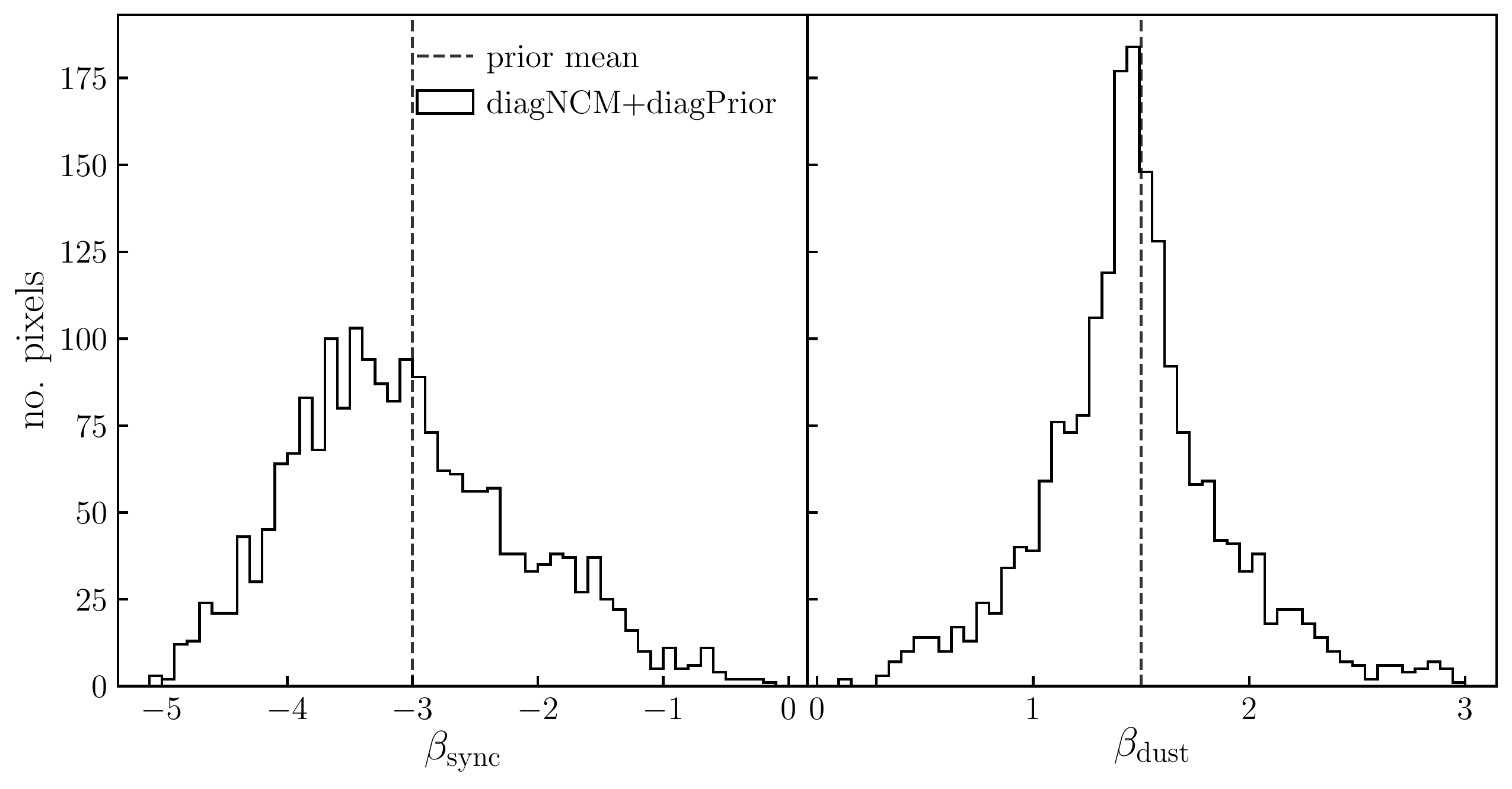}
    \vspace{-0.1in}
    \caption[]{Histograms of spectral index maps shown in Fig.~\ref{fig:pix_mask_beta} using a mask with $f_{\rm sky}=0.66$. The input priors are shown as dashed vertical black lines for both spectral indices.}
    \label{fig:hist_mask_beta}
    \vspace{-0.1in}
\end{figure}

In Fig.~\ref{fig:pix_mask_beta} we show the recovered spectral index maps for $\bsync$ and $\bdust$ with the corresponding histograms shown in Fig.~\ref{fig:hist_mask_beta}. The recovered mean and standard deviation for the diagNCM+diagPrior case with $f_{\rm sky}=0.66$ is
\begin{subequations}
\begin{align}
\overline{\beta}_{\rm sync} = -3.055\pm 0.916\ , \label{eq:beta_s_fsky_066_1} \\
\overline{\beta}_{\rm dust} = \phantom{-}1.470 \pm  0.439\ ,  \label{eq:beta_s_fsky_066_2}
\end{align}
\end{subequations}
and are consistent with the corresponding \SRone diagNCM+diagPrior result presented in Table \ref{tab:beta_distribution}. We note that the recovered scatter in the spectral index maps is larger but attribute this to the simplified noise model used for this test. 

Corresponding to the results discussed in Sec.~\ref{sec:foreground_res}, we find evidence for synchrotron spectral index variations in the radio loops I, III, IIIb, VIIb and IV, given in Table \ref{tab:beta_distribution_loops_mask}.

\begin{table}
\centering
{\def\arraystretch{1.2}\tabcolsep=5pt
\begin{tabular}{llcc}
\hline \hline
method    & loop & $\overline{\beta}_{\rm sync}$      & $\overline{\beta}_{\rm dust}$  \\ \hline
diagNCM + diagPrior & I &$-3.502 \pm 0.675$& $1.351 \pm 0.245$\\
                    & III &$-3.889 \pm 0.735$& $1.746 \pm 0.366$\\
                    & IIIs  &$-3.473 \pm 0.527$& $1.462 \pm 0.269$\\ 
                    & VIIb  &$-3.568 \pm 0.637$& $1.541 \pm 0.432$\\ 
                    & XII  &$-2.941 \pm 0.609$& $1.559 \pm 0.466$\\ 
                    & IV  &$-3.614 \pm 0.741$& $1.176 \pm 0.233$\\ \hline
\end{tabular}}
\caption{Summary of spectral index distributions for synchrotron and dust in six loops, shown in Fig.~\ref{fig:mask_loop}, for \SRone data using a mask with $f_{\rm sky}=0.66$.}
\label{tab:beta_distribution_loops_mask}
\vspace{-0.1in}
\end{table}

\section{Testing for latitudinal spectral index variations}\label{app:lat_dependence}
\begin{figure}
    \centering
     \includegraphics[width=0.8\columnwidth]{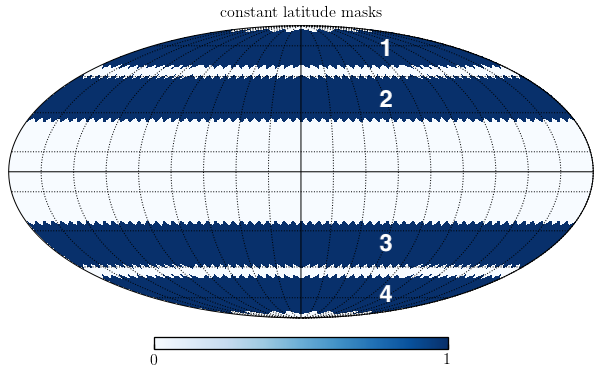}
    \vspace{-0.1in}
    \caption[]{Four stripes in Galactic latitude, labelled 1 to 4, used to investigate potential spectral index variation. Each stripe has a width of 13$^\circ$ in latitude and they are centred at $\pm38^\circ$ (stripes 2 and 3) and $\pm70^\circ$ (stripes 1 and 4).}
    \label{fig:mask_lat}
    \vspace{-0.1in}
\end{figure}

In this appendix, we test for spectral index variations as a function of Galactic latitude using \SRtwo data for HFI. We investigate four disjoint regions, shown in Fig.~\ref{fig:mask_lat}. Two `stripes' are located in the northern and two stripes in the southern hemisphere, avoiding the Galactic plane and the Galactic poles. For the three pixel-correlation cases we consider, we obtain the results quoted in Table \ref{tab:beta_distribution_lat_mask}. We do not see any significant spectral index variations in either dust or synchrotron between these stripes.

\begin{table}
\centering
{\def\arraystretch{1.2}\tabcolsep=5pt
\begin{tabular}{lccc}
\hline \hline
case & stripe     & $\overline{\beta}_{\rm sync}$      & $\overline{\beta}_{\rm dust}$  \\ \hline
corrNCM+corrPrior&1 & $-2.759 \pm 0.642$  &$1.423 \pm 0.273$\\
                 &2 & $-2.846 \pm 0.625$  &$1.472 \pm 0.248 $\\
                 &3 & $-2.882 \pm 0.596$  &$1.392 \pm 0.199$\\
                 &4 & $-2.860 \pm 0.582$  &$1.384 \pm 0.229 $\\
corrNCM+diagPrior&1 & $-2.875 \pm 0.533$  &$1.444 \pm 0.233$\\
                 &2 & $-2.893 \pm 0.550$  &$1.482 \pm 0.214$\\
                 &3 & $-2.928 \pm 0.537$  &$1.414 \pm 0.191$\\
                 &4 & $-2.947 \pm 0.471$  &$1.433 \pm 0.196$\\
diagNCM+diagPrior&1 & $-2.769 \pm 0.939$  &$1.343 \pm 0.480$\\
                 &2 & $-2.939 \pm 0.906$  &$1.509 \pm 0.429$\\
                 &3 & $-3.253 \pm 0.701$  &$1.471 \pm 0.311$\\
                 &4 & $-2.954 \pm 0.743$  &$1.406 \pm 0.375$\\ \hline
\end{tabular}}
\caption{Summary of spectral index distributions for synchrotron and dust in four stripes in Galactic latitude, shown in Fig.~\ref{fig:mask_lat}, for \SRtwo data.}
\label{tab:beta_distribution_lat_mask}
\vspace{-0.1in}
\end{table}

\bsp	
\label{lastpage}
\end{document}